\documentclass[pra,twocolumn,superscriptaddress,showpacs,floatfix,amssymb,
amsmath,fontsize=11pt]{revtex4-1}

\usepackage{times}
\usepackage{amssymb}
\usepackage[pdftex]{graphicx}
\usepackage{dcolumn}
\usepackage[latin1]{inputenc}
\usepackage[mathscr]{eucal}
\usepackage{bm}
\usepackage{bbm}
\usepackage{textcomp}
\usepackage{sidecap}
\usepackage{times}
\usepackage[american]{babel}

\newcommand{\ho}{\tilde{h}_0^{}}
\newcommand{\ebk}{\varepsilon_{\bm{k}}^{}}
\newcommand{\ela}{\varepsilon_{\Lambda}^{}}
\newcommand{\elaa}{\varepsilon_{\Lambda}^{2}}
\newcommand{\nbe}{n_{\scriptscriptstyle\mathrm{BE}}^{}}
\newcommand{\mf}{m_{\scriptscriptstyle{F}}^{}}
\newcommand{\be}{\begin{equation}}
\newcommand{\ee}{\end{equation}}
\newcommand{\bee}{\begin{equation*}}
\newcommand{\eee}{\end{equation*}}
\newcommand{\bx}{\bm{x}}
\newcommand{\un}{\hat{\bm{n}}}
\newcommand{\bq}{\bm{q}}
\newcommand{\br}{\bm{r}}

\newcommand{\bk}{\bm{k}}
\newcommand{\dk}{\mathrm{d}\bm{k}}

\newcommand{\la}{\langle}
\newcommand{\ra}{\rangle}
\newcommand{\omn}{\omega_n^{}}
\newcommand{\al}{\alpha}
\newcommand{\bt}{\beta}
\newcommand{\rd}{\mathrm{d}}
\newcommand{\gm}{\gamma}

\newcommand{\kb}{k_{\scriptscriptstyle\mathrm{B}}}

\begin{document}

\title{Phase transitions in dipolar spin-$1$ Bose gases}

\author{Ville Pietil\"a}
\affiliation{Department of Applied Physics/COMP, Aalto
  University, P.~O.~Box 14100, FI-00076 AALTO, Finland}
\affiliation{Department of Physics, Harvard University, Cambridge, 
Massachusetts 02138, USA}

\author{Mikko M\"ott\"onen}
\affiliation{Department of Applied Physics/COMP, Aalto
  University, P.~O.~Box 14100, FI-00076 AALTO, Finland}
\affiliation{Low Temperature Laboratory,
Aalto University, P.~O.~Box 13500, FI-00076 AALTO,
Finland}

\begin{abstract}
We study phase transitions in homogeneous spin-$1$ Bose gases in the presence 
of long-range magnetic dipole--dipole interactions (DDI). We concentrate on  
three-dimensional geometries and employ momentum shell renormalization group to 
study the possible instabilities caused by the dipole--dipole interaction. At 
the zero-temperature limit where quantum fluctuations prevail, we find the 
phase diagram to be unaffected by the dipole--dipole interaction. When the thermal fluctuations 
dominate, polar and ferromagnetic condensates with DDI become unstable and we discuss 
this crossover in detail.  On the other hand, the spin-singlet condensate remains stable in the 
presence of  DDI.
\end{abstract} 


\maketitle

\section{Introduction} 

The past few years have revealed that ultracold atomic gases can answer 
important questions beyond the immediate scope of atomic 
physics~\cite{Micheli:2006,Donner:2007,Jordens:2008,Jo:2009}. In particular, 
experimental methods have matured to the level where measurements of 
critical exponents are possible in some cases~\cite{Donner:2007}. This provides 
an interesting opportunity to study the physics of phase transitions and critical 
phenomena utilizing 
cold atomic gases, as well as to realize exotic phases that are absent in more  
conventional solid state systems~\cite{Gorshkov:2010}. In this work, we consider 
Bose gases with a spin degree of freedom~\cite{Ho:1998,Ohmi:1998}. They provide an 
intriguing example where magnetic ordering can compete with superfluidity and 
condensation. This interplay can give rise to a myriad of topological 
defects~\cite{Leonhardt:2000,Stoof:2001,AlKhawaja:2001,Savage:2003,Takahashi:2007,
Kawaguchi:2008,Pietila:2009,Huhtamaki:2010,Huhtamaki:2010b} which play an important role, 
e.g., in the superfluid transition in low-dimensional systems~\cite{Mukerjee:2006,Pietila:2010}.

Initially, the magnetic properties of spinor Bose gases were assumed to depend only 
on the local interactions determined by the scattering lengths in the different 
total hyperfine spin channels~\cite{Ho:1998,Ohmi:1998,
Stenger:1998,Schmaljohann:2004,Chang:2004,Sadler:2006}, but recent 
experiments suggest that the long-range magnetic dipole--dipole interaction (DDI) may be   
an essential ingredient in determining the properties of spinor Bose 
gases~\cite{Griesmaier:2005,Lahaye:2007,Koch:2008,Vengalattore:2008,Vengalattore:2010}. 
In this work, we consider the  
effect of dipole--dipole interaction in spin-$1$ Bose gases using momentum shell  
renormalization group (RG)~\cite{Wilson:1974,Shankar:1994,Fisher:1988,Kolomeisky:1992,Kolomeisky:1992b,Bijlsma:1996}. We note that also the functional renormalization 
group~\cite{Wetterich:1991} has been successfully applied in the context of cold 
atoms~\cite{Andersen:1999,Andersen:2004,Diehl:2010}. The momentum shell RG analysis 
allows us to determine the effect of DDI on the phase diagram of spin-$1$ Bose gases which has 
recently attracted some interest~\cite{Yang:2009,Kolezhuk:2010,Natu:2011}. 
Moreover, the recent advances in the creation of Feshbach resonances using either optical 
means~\cite{Hamley:2009} or microwaves~\cite{Papoular:2010}, suggest that exploration of 
the phase diagram could become experimentally realistic in the near future.

Dipole--dipole interaction couples the spin directly to spatial degrees 
of freedoms, giving local spins tendency to align head to tail and antialign 
side by side~\cite{Cherng:2009,Lahaye:2009}. On the other hand, the experiments 
described in Refs.~\cite{Vengalattore:2008,Vengalattore:2010} are of mixed dimensionality 
in the sense that spin dynamics was effectively two-dimensional while otherwise the 
system was spatially three-dimensional (3D). Furthermore, the original DDI was   
strongly modified by a rapid Larmor precession induced by an external magnetic field. 
In the present work, we focus on the properties of pristine DDI and consider 
a homogeneous three-dimensional system in the absence of external magnetic fields. 
For three-dimensional systems, DDI is a true long-range 
interaction~\cite{Astrakharchik:2008,Lahaye:2009} and we avoid additional complications 
that may arise due to the absence of true long-range order in low-dimensional systems.

Although dipole--dipole interactions are present in all ferromagnetic materials, 
they are usually weak and often neglected or treated 
phenomenologically~\cite{Ma:1976,Snoke:2008}. However, for 
ferromagnets which order only at very low temperatures, DDI might be  
crucial for the correct low-energy behavior~\cite{Belitz:2010}. In this work, we analyze this 
scenario in the context of spin-$1$ Bose gases. We find that DDI introduces additional 
instabilities to the expected finite-temperature phase diagram~\cite{Yang:2009}. 
In particular, DDI renders both polar and ferromagnetic condensates unstable and the RG analysis alludes to the existence of a fluctuation-induced first-order 
transition. 

In the zero-temperature limit, we show that DDI  renormalizes to zero and the usual mean-field 
theory~\cite{Ho:1998,Ohmi:1998,Lahaye:2009} is a valid description of the 
system. Dipole--dipole interactions also generate a new single-particle term which 
has not been taken into account in the previous studies. This new interaction is 
relevant in the RG sense and it is allowed by the symmetries of the 
system. However, in the zero-temperature limit it renormalizes to zero along with 
the DDI. 

\section{The model \label{model}}

We consider a uniform spin-$1$ Bose gas neglecting the effects 
of an external potential that confines the atoms. In the presence of DDI, the system  
has a global $U(1)$ symmetry associated with the conserved atom number and a global $SO(3)$ 
symmetry corresponding to a simultaneous rotation of spin and coordinate 
spaces~\cite{Yi:2006,Kawaguchi:2006}. The latter symmetry indicates that only the sum 
of spin and orbital angular momentum is conserved. The effective action in the Zeeman 
basis $\{|F=1,\mf=+1,0,-1\ra\}$ can be written as $S = S_0^{} + S_{\mathrm{int}}^{}$, 
\begin{widetext}
\begin{align}
\label{s_0}
&S_0^{} = \int_{0}^{\hbar\bt}\mathrm{d}\tau\int\mathrm{d}\bx\,\,\psi_a^*(\bx,\tau)
\big(\hbar\Gamma^{-1}_{}\partial_{\tau}^{} - \frac{\hbar^2}{2m}\nabla^2-\mu\big)
\psi_a^{}(\bx,\tau),  \\
\label{s_int}
&S_{\mathrm{int}}^{} = \int_{0}^{\hbar\bt}\mathrm{d}\tau\int\mathrm{d}\bx\,\mathrm{d}\bx'\,
\bigg[\frac{c_0^{}}{2}\,|\psi_a^{}(\bx,\tau)|^4\delta(\bx-\bx') + \frac{c_2^{}}{2}\,
|\bm{\mathcal{S}}(\bx,\tau)|^2\delta(\bx-\bx') + \notag \\ 
& \hspace{2cm} \frac{c_{\mathrm{dd}}^{}}{2}\,\mathcal{S}^i_{}(\bx,\tau)h_{}^{ij}(\bx-\bx')
\mathcal{S}^j_{}(\bx',\tau)\bigg],
\end{align}
\end{widetext}
where $\tau$ is the imaginary time and $\beta = 1/\kb T$.  We always assume implicit summation over repeated indices. The local spin is given by 
$\mathcal{S}^i = \psi_{a}^*\,\mathcal{F}_{ab}^i\,\psi_{b}^{}$, where $\mathcal{F}_{}^i$ 
are spin-$1$ matrices in the Zeeman basis and $\psi_a^{}$, $a=-1,0,1$ are bosonic fields. 
Parameter $\Gamma$ is initially set to unity and it acquires nontrivial renormalization under the 
RG transformation. In this work we consider two distinct limits: the $T=0$ case where 
$\Gamma$ renormalizes only due to the anomalous dimension of the fields $\psi_a^{}$ and 
the high temperature limit where $\Gamma$ renormalizes to zero and we obtain a classical 
theory.

\begin{figure}[b!]
\begin{center}\includegraphics[width=0.425\textwidth]{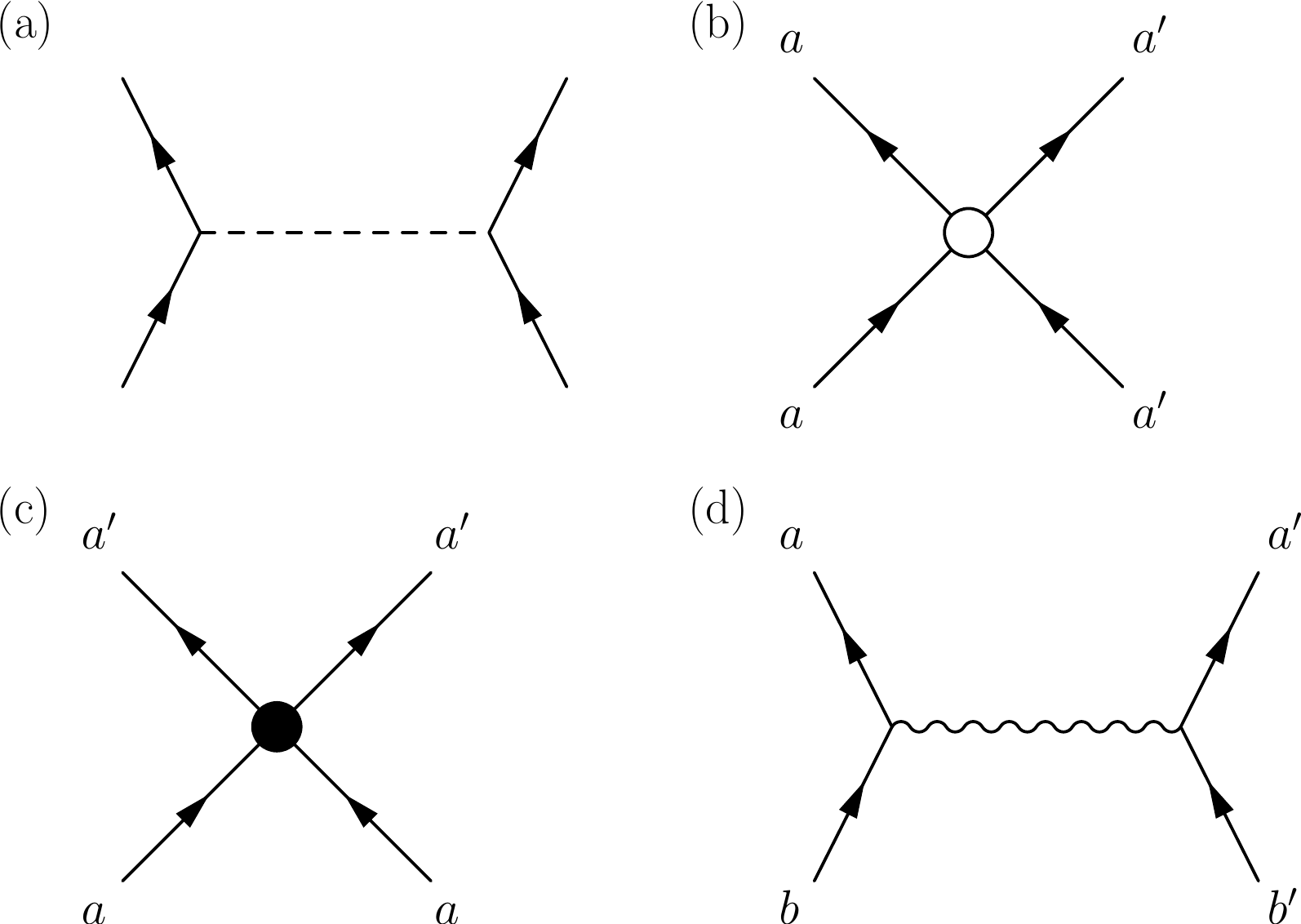}
\caption{\label{vertices}(a)~Generic interaction vertex, (b)~local density--density interaction, 
(c)~local spin-spin interaction, (d)~and dipole--dipole interaction. The generic interaction 
vertex (a) can denote any of the vertices (b)--(d). Conservation of the 4-momentum 
$\vec{k}=(\omn,\bk)$ at each vertex is implied.} 
\end{center}
\end{figure}

The bare values of the coupling constants $c_0^{}$ and $c_2^{}$ are related to 
scattering lengths $a_0^{}$ and $a_2^{}$ in the total hyperfine spin channels $F=0$ and 
$F=2$ by $c_0^{}=4\pi\hbar^2(a_0^{} + 2a_2^{})/3m$ and $c_2^{}=4\pi\hbar^2(a_2^{} - 
a_0^{})/3m$. The coupling constant corresponding to DDI is given by  
$c_{\mathrm{dd}}^{} = \mu_0^{}\mu_{\scriptscriptstyle\mathrm{B}}^{2}
g_{\scriptscriptstyle\mathrm{F}}^{2}/4\pi$,  where $\mu_0^{}$ is the vacuum permeability, 
$\mu_{\scriptscriptstyle\mathrm{B}}^{}$ Bohr magneton, and  $g_{\scriptscriptstyle\mathrm{F}}$  
Land\'e $g$-factor. The kernel for dipole--dipole interactions in the momentum space takes 
the form~\cite{Cherng:2009}
\be
\label{dipole_kernel}
h_{}^{ij}(\bq) = -\frac{4\pi}{3}(\delta_{}^{ij} - 3\hat{q}_{}^{i}\hat{q}_{}^{j}).
\ee
In real space, $h_{}^{ij}(\br)$ decays as $1/|\br|^3$.
To date, the only experimentally studied dipolar spin-$1$  Bose gas has been 
$^{87}$Rb for which the different coupling constants satisfy $c_2^{}/c_0^{} = -0.005$ and 
$c_{\mathrm{dd}}^{}/|c_2^{}| = 0.1$~\cite{Vengalattore:2008}. Another candidate for 
dipolar spin-$1$ Bose gas is $^{23}$Na for which the scattering lengths of 
Ref.~\cite{Burke:1998} give $c_2^{}/c_0^{} = 0.03$ and 
$c_{\mathrm{dd}}^{}/c_2^{} = 0.006$. The small value of $c_{\mathrm{dd}}^{}/c_2^{}$ explains 
why the effects arising from DDI have been expected to be vanishingly small for $^{23}$Na. 
The different interaction vertices appearing in the RG calculations in  
Sections~\ref{rg}--\ref{ddi_rg_full} are illustrated in Fig.~\ref{vertices}.

To streamline the RG calculations, we switch to  the Cartesian basis. 
In this basis, the field operator $\Psi = (\psi_x,\psi_y,\psi_z)$ transforms as a vector under 
spin rotations. Moreover, the spin-1 matrices $(\mathcal{F}_x,\mathcal{F}_y,\mathcal{F}_z)$ 
take a particularly simple form $(\mathcal{F}_\al)_{\beta\gamma}^{} = -i\varepsilon_{\al\bt\gm}$,  
where $\varepsilon_{\al\bt\gm}$ is the Levi-Civita tensor. The effective action can be written as 
\begin{align}
\label{s0}
S_0^{} &= \int \mathrm{d}^4k\,\,\psi_\al^*(\vec{k})\big(-i\hbar\Gamma^{-1}_{}\omn + 
\varepsilon_{\bk}^{}-\mu\big)\psi_\al^{}(\vec{k}), \\
\label{sint}
S_{\mathrm{int}}^{} &= \int\,\frac{\mathrm{d}^4k\,\mathrm{d}^4p\,\mathrm{d}^4q}{(2\pi)^3
\hbar\beta}\,\bigg[\frac{g_d^{}}{2}\,\psi_\al^*(\vec{p}+\vec{q})\psi_{\al'}^*(\vec{k}-\vec{q})
\psi_{\al'}^{}(\vec{k})\psi_{\al}^{}(\vec{p}) \notag \\
& \hspace{3mm} +\,\frac{g_s^{}}{2}\,\psi_\al^*(\vec{p}+\vec{q})\psi_{\al}^*(\vec{k}-\vec{q})
\psi_{\al'}^{}(\vec{k})\psi_{\al'}^{}(\vec{p}) \notag \\ 
+ &\,\frac{c_{\mathrm{dd}}^{}}{2}\,\psi_\al^*(\vec{p}+\vec{q})\psi_{\al'}^*(\vec{k}-\vec{q})\,h_{}^{ij}(\bq)
\mathcal{F}_{\al\bt}^i\mathcal{F}_{\al'\bt'}^j\,\psi_{\bt'}^{}(\vec{k})\psi_{\bt}^{}(\vec{p})\bigg],
\end{align}
where indices $\{x,y,z\}$ are referred to by the Greek indices $\al,\bt,...$ and the 
Latin indices $a,b,...$ correspond to the original Zeeman basis. 
We use a shorthand notation $\vec{k} = (\omn,\bk)$ and  
$\int\mathrm{d}^4k = \sum_{\omn}\int\dk$.  Bosonic Matsubara frequencies are given by 
$\omn = 2\pi n/\hbar\beta$ and $\varepsilon_{\bk}^{} = \hbar^2\bk^2/2m$. The coupling 
constants $g_d^{}$ and $g_s^{}$ are related to the coupling constants in the Zeeman 
basis by $g_d^{} = c_0^{} + c_2^{}$ and $g_s^{}=-c_2^{}$.

\section{\label{rg}Renormalization group calculation}

We set up the RG calculation in a fixed dimension $D$ following  
Refs.~\cite{Fisher:1988,Kolomeisky:1992,Kolomeisky:1992b,Bijlsma:1996}. 
To make a connection to Refs.~\cite{Yang:2009,Kolezhuk:2010}, we first neglect the 
dipole--dipole interactions and study a general $D$-dimensional situation. We show that 
our RG equations at the zero-temperature limit coincide with   
Ref.~\cite{Kolezhuk:2010} and essentially reproduce the phase diagram proposed in 
Ref.~\cite{Yang:2009}. We also point out that in a contrast to  
Ref.~\cite{Kolezhuk:2010}, in which the stability of low-dimensional multi-component Bose 
gases was considered at the zero-temperature limit, our main focus is a three 
dimensional spinor Bose gas at finite temperatures. 
We note that isotropic long-range interactions in spinless Bose gases have been analyzed in 
the zero-temperature limit in Ref.~\cite{Kolomeisky:1992b} and long-range interactions of the form $V(\br)\propto1/|\br|^s$ were found irrelevant for $s>2$. 
Our findings in the presence of DDI are similar to those of Ref.~\cite{Kolomeisky:1992b}, 
namely, DDI becomes irrelevant at zero temperature.

To study the effects of DDI, we employ the momentum shell 
RG~\cite{Wilson:1974,Shankar:1994,Bijlsma:1996} in which 
we split the fields appearing in Eqs.~\eqref{s0} and~\eqref{sint} such that 
$\psi_\al^{} = \psi_{\al,<}^{} + \psi_{\al,>}^{}$, where the $\psi_{\al,<}^{}$ contains 
momentum components with $|\bk| < \Lambda/s$ and $\psi_{\al,>}^{}$ corresponds to momenta 
$\Lambda/s \leq |\bk| < \Lambda$. The ultraviolet (UV) cutoff is denoted by $\Lambda$, and 
in general, nonuniversal quantities such as condensate fraction or critical 
temperature depend explicitly on $\Lambda$. Several tricks such as halting the RG flow 
when an appropriate scale is reached or relating $\Lambda$ to the $s$-wave scattering 
length can be used to obtain information on quantities depending on 
$\Lambda$~\cite{Kolezhuk:2010,Fisher:1988,Kolomeisky:1992,Kolomeisky:1992b,Bijlsma:1996}.

\begin{figure}[h!]
\begin{center}
\includegraphics[width=0.425\textwidth]{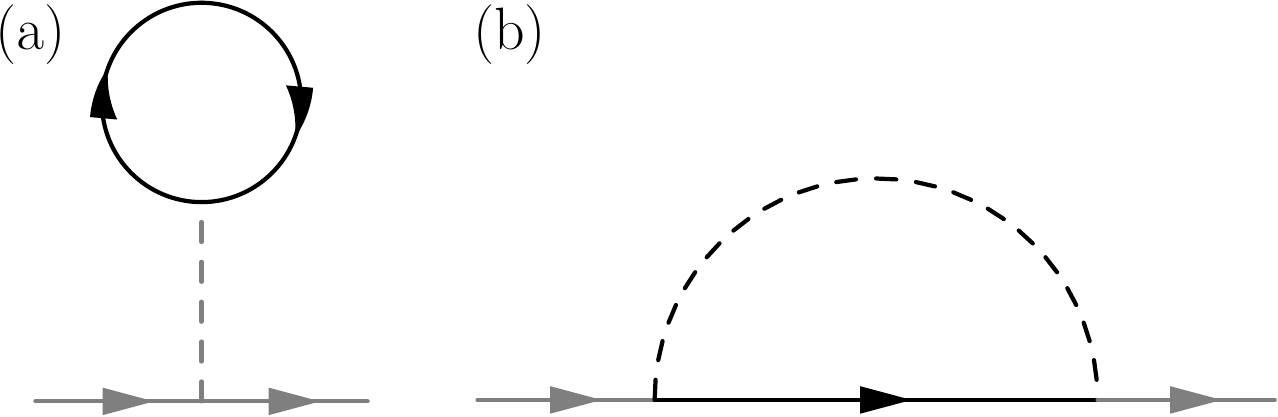}
\caption{One-loop diagrams contributing to renormalization of the chemical potential. 
The interaction line can be any of those denoted in Fig.~1(b)--(d). The external legs 
corresponding to fields $\psi_{\al,<}^{}$ and $\psi_{\al,<}^{*}$ are denoted by gray lines 
for clarity. The dipole--dipole interaction can give rise to external interaction lines which are 
also denoted by gray dashed lines [the tadpole diagram in (a)].} 
\label{chempot}
\end{center}
\end{figure}

\begin{figure}[h!]
\begin{center}
\includegraphics[width=0.425\textwidth]{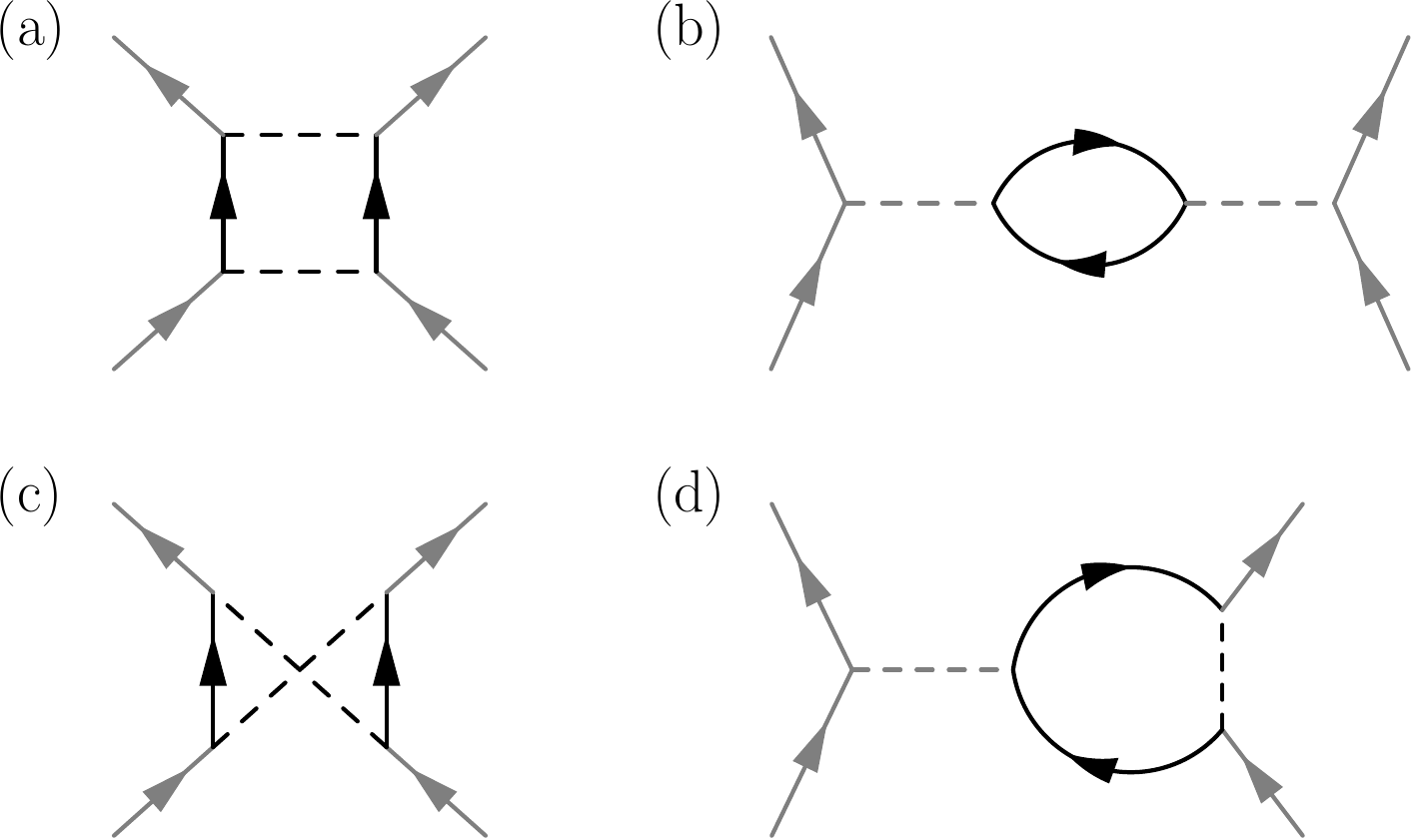}
\caption{\label{couplings}One-loop diagrams contributing to renormalization of coupling 
constants $g_d^{}$, $g_s^{}$, and $c_{\mathrm{dd}}^{}$. The two interaction lines can 
be any combination of the interactions in Fig.~1(b)--(d). The external legs and interaction 
lines are again denoted by gray color.}
\end{center}
\end{figure}

The RG calculation proceeds by integrating out the fast modes residing at the momentum 
shell $\Lambda/s \leq |\bk| < \Lambda$, which results in a renormalized action for the 
slow modes $\psi_{\al,<}^{}$. At the second step of RG transformation, the UV cutoff is 
brought back to the original value $\Lambda$ by rescaling the fields and momenta, 
giving rise to RG equations for the chemical potential and coupling constants. Only the 
one-particle irreducible connected diagrams contribute to the RG equations. In this work, 
we compute the RG equations to the one-loop order. The relevant diagrams appearing in 
the renormalization of chemical potential and coupling constants are shown 
Figs.~\ref{chempot} and~\ref{couplings}. 

The diagrams in Figs.~\ref{chempot} and~\ref{couplings} correspond to an expansion with 
respect to coupling constants $g_d^{}$, $g_s^{}$, and $c_{\mathrm{dd}}^{}$ in the first 
non-trivial order. The internal lines are evaluated using non-interacting 
one-particle Green's function
\be
\label{plain_propagator}
\mathcal{G}_{0,\al\bt}^{}(\bk,\omn) = -\frac{\hbar}{-i\hbar\Gamma^{-1}_{}\omn + 
\varepsilon_{\bk}^{} - \mu}\delta_{\al\bt}^{}.
\ee
After integrating out the fast modes and neglecting the irrelevant terms generated by 
the momentum shell integration, the slow fields, momentum, and imaginary time are 
rescaled as~\cite{Fisher:1988}
\begin{subequations}
\label{scaling1}
\begin{align}
& \bk \rightarrow \bk\,e^{-\ell}_{}, \\
& \tau \rightarrow \tau\,e^{z\ell}_{}, \\
& \psi_\al^{} \rightarrow \psi_\al^{}\,e^{\zeta\ell}_{},
\end{align}
\end{subequations}
where we have taken $s=e^\ell$. For simplicity, we first neglect the anomalous dimension of 
the fields, which allows us to keep the kinetic energy term in Eq.~\eqref{s0} fixed during the RG 
transformation. In Section~\ref{ddi_rg_full}, we take into account also the 
renormalization of the kinetic energy term and find that $\varepsilon_{\bk}^{}$ is only weakly 
renormalized. For vanishing anomalous dimension we obtain an identity~\cite{Fisher:1988}
\be
\label{scaling_identity}
2\zeta + z = 2-D,
\ee
and the relevance of all other terms is compared to the kinetic energy. The requirement 
that the rescaled action is equivalent to the original one yields the scaling relations
\begin{subequations}
\label{scaling2}
\begin{align}
\label{gamma_scaling}
&\Gamma \rightarrow \Gamma\,e^{(d+2\zeta)\ell}_{},\\
\label{mu_scaling}
&\mu \rightarrow \mu\,e^{-2\ell}_{}, \\
\label{g_scaling}
&g \rightarrow g\,e^{-2(\zeta+1)\ell}_{},\\
&T \rightarrow T\,e^{-z\ell}
\end{align}
\end{subequations}
where $g = g_d^{}, g_s^{},c_{\mathrm{dd}}^{}$, and we have used Eq.~\eqref{scaling_identity}. 
In the presence of DDI, we take $D=3$. 


In general, RG calculations provide information on universal quantities such as 
different phases and transitions between them. On the other hand, renormalization group 
analysis can be used to determine the relevance of a particular interaction for a given phase 
and to study the stability of different phases when additional interactions are included. We 
take the latter point of view in Sections~\ref{no_ddi}--\ref{ddi_rg_full} where we analyze the 
effects of DDI on the phase diagram of spin-$1$ Bose gases.

\section{\label{no_ddi}RG flow in the absence of dipole--dipole interactions}

\begin{figure}[h!]
\begin{center}
\includegraphics[width=0.375\textwidth]{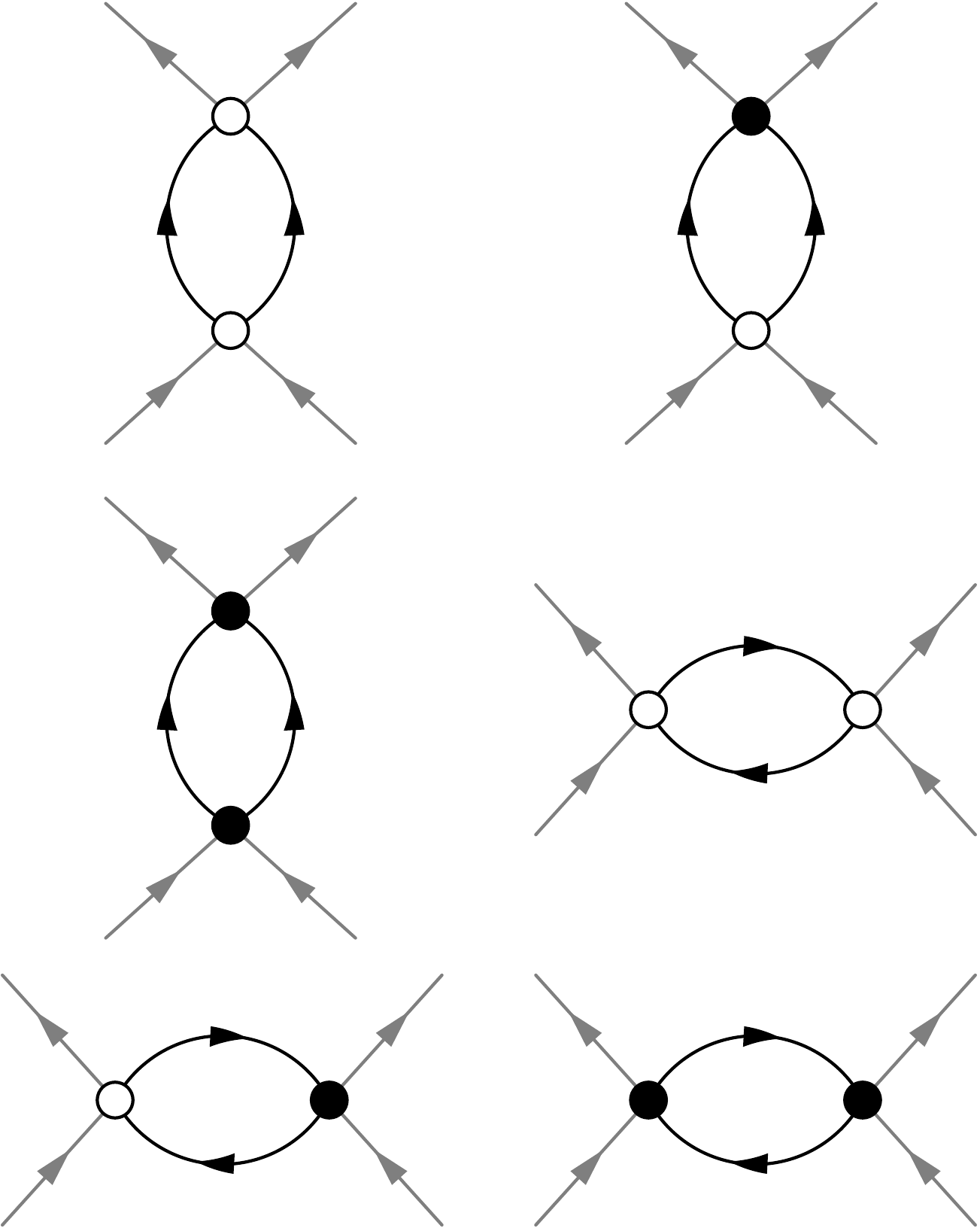}
\caption{\label{no_ddi_diagrams}One-loop contributions to RG equations in the absence of 
dipole--dipole interactions.}
\end{center}
\end{figure}

\begin{figure}[t!]
\begin{center}
\includegraphics[width=0.425\textwidth]{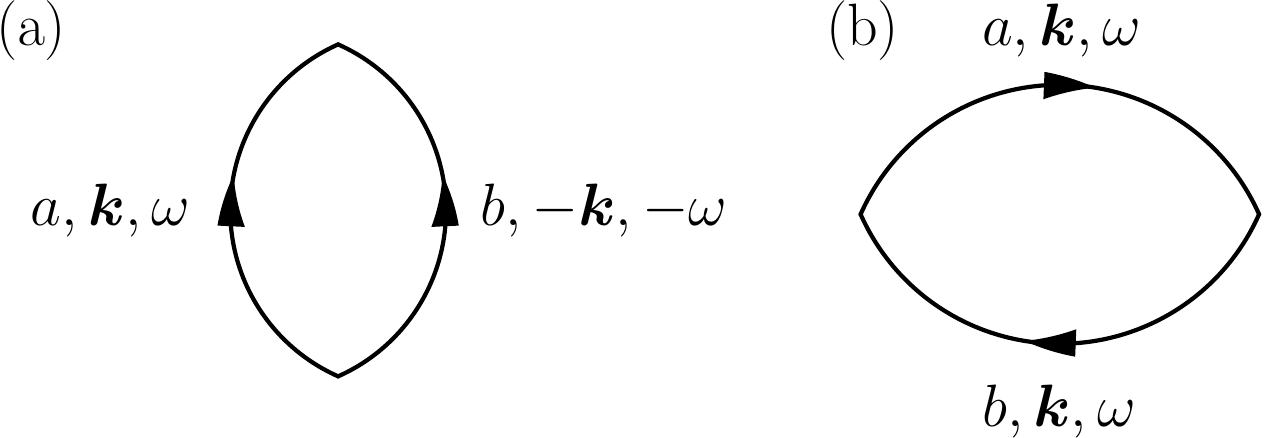}
\caption{\label{bubbles} Basic bubbles for diagrams in the absence of dipole--dipole 
interactions. The bubbles depend in principle on external momenta. Expanding about 
the zero momentum one observes that terms corresponding to finite external momentum 
are at most marginal in the RG sense. Since we have neglected 
marginal interactions from the outset, we take the external momenta to be zero.}
\end{center}
\end{figure}

Let us first consider a general dimension $D$ and neglect DDI. The different 
diagrams contributing to the RG equations are shown in Fig.~\ref{no_ddi_diagrams}.
Since all interactions are local, the different diagrams in Fig.~\ref{no_ddi_diagrams} 
reduce to evaluation of the bubbles shown in Fig.~\ref{bubbles}. Integration on the 
momentum shell is restricted to the interval $\Lambda/s \leq |\bk| < \Lambda$ and the 
Matsubara sums can be calculated using the standard methods~\cite{Bruus:2004}. At the 
limit of an infinitesimal shell of thickness $\Lambda d\ell$, we obtain
\begin{align}
&\rd\mu = -2(g_s^{} + 2g_d^{})K_D^{}\Gamma\,\nbe[\bt\Gamma(\ela-\mu)]\Lambda^D \rd\ell, \\
&\rd g_s^{} = -[(3g_s^{2} + 2g_d^{}g_s^{})\chi_1^{} + 4g_d^{}g_s^{}\chi_2^{}]K_D^{}
\Lambda^D\rd\ell, \\
&\rd g_d^{} = -[g_d^2\chi_1^{} + 4(g_d^2 + g_d^{}g_s^{} + g_s^2)\chi_2^{}]K_D^{}\Lambda^D 
\rd\ell,
\end{align}
where $K_D^{} = [2^{D-1}\pi^{D/2}\Gamma(D/2)]^{-1}$ and $\Gamma(x)$ is the Gamma function 
(not to be confused with the energy parameter $\Gamma$).  
Furthermore, functions $\chi_1^{}(\bt,\Gamma)$ and $\chi_2^{}(\bt,\Gamma)$ are given by  
\begin{align}
&\chi_1^{}(\bt,\Gamma) = \Gamma\,\frac{1+2\nbe[\bt\Gamma(\ela-\mu)]}
{2(\ela-\mu)}, \\
&\chi_2^{}(\bt,\Gamma) = \bt\Gamma^2\,\nbe[\bt\Gamma(\ela-\mu)]\{1+\nbe[\bt
\Gamma(\ela-\mu)]\}.
\end{align}
We have denoted the Bose distribution function by $\nbe(x) = 1/(e^x-1)$ and  
$\ela = \hbar^2\Lambda^2/2m$. The contributions 
proportional to $\chi_1^{}$ correspond to diagrams containing the bubble in 
Fig.~\ref{bubbles}(a) and contributions containing $\chi_2^{}$ arise from diagrams with 
the bubble in Fig.~\ref{bubbles}(b).

\vspace{5mm}

Taking into account the scalings dictated by Eqs.~\eqref{scaling1} 
and~\eqref{scaling2}, using the scaling relation in Eq.~\eqref{scaling_identity}, and  
transforming back to the Zeeman basis we obtain the following flow equations
\begin{align}
\label{beta_rg}
&\frac{\rd\bt}{\rd\ell} = -z\bt, \\
\label{gamma}
&\frac{\rd\Gamma}{\rd\ell} = -(2\zeta+D)\Gamma, \\
\label{mu_rg}
&\frac{\rd\mu}{\rd\ell} = 2\mu -2(c_2^{}+2c_0^{})K_D^{}\Lambda^D\Gamma\,
\nbe[\bt\Gamma(\ela-\mu)], \\
\label{c0_rg}
&\frac{\rd c_0^{}}{\rd\ell} = 2(\zeta+1)c_0^{} - [(c_0^2+2c_2^2)\chi_1^{} + 
4c_0^2\chi_2^{}]K_D^{}\Lambda^D, \\
\label{c2_rg}
&\frac{\rd c_2^{}}{\rd\ell} = 2(\zeta+1)c_2^{} -[(2c_0^{}c_2^{} - c_2^2)\chi_1^{} \notag \\ 
&\hspace{3.7cm} + 4(c_0^{}c_2^{} + c_2^2)\chi_2^{}]K_D^{}\Lambda^D. 
\end{align}
Although these equation are valid for any temperature, we consider two limits:  
the quantum regime which takes place at the zero-temperature limit and is dominated by 
the quantum fluctuations, and the thermal regime where thermal fluctuations prevail 
over the quantum fluctuations~\cite{Fisher:1988}.

\subsection{\label{no_ddi_quantum}Quantum regime}

Let us first consider the quantum regime at which we require $\rd\Gamma/\rd\ell=0$. Furthermore, 
we set $\Gamma=1$. From Eqs.~\eqref{scaling_identity} and~\eqref{gamma} we 
obtain $\zeta = -D/2$ and $z=2$. This gives the usual instability of the $T^*=0$ fixed 
point, since any nonzero temperature tends to increase in the RG flow. In the limit 
$\kb T \ll \ela - \mu$, we have 
\begin{align*}
&\chi_1^{}(\bt,\Gamma) \rightarrow \frac{1}{2(\ela-\mu)}, \\
&\chi_2^{}(\bt,\Gamma) \rightarrow 0,
\end{align*}
where the latter equation gives the well-known result stating that in the zero-temperature limit,  
only the ladder diagrams contribute to the renormalization~\cite{Sachdev:1999}. The RG 
equation for the chemical potential becomes $\rd\mu/\rd\ell = 2\mu$, and the only 
fixed point is $\mu^*=0$. Furthermore, the remaining RG equations reduce to
\begin{align}
\label{c0_first}
&\frac{dc_0^{}}{d\ell} = (2-D)c_0^{} - (c_0^2 + 2c_2^2)\frac{K_D^{}\Lambda^D}
{2\ela}, \\
\label{c2_first}
&\frac{dc_2^{}}{d\ell} = (2-D)c_2^{} + (c_2^2 - 2c_0^{}c_2^{})\frac{K_D^{}\Lambda^D}
{2\ela}.
\end{align}
The above equations are precisely those of Ref.~\cite{Kolezhuk:2010}, and for the future 
reference, we consider here the $D=3$ case. Cases $D=1$ and $D=2$ have been 
analysed in Ref.~\cite{Kolezhuk:2010}.

\begin{table}[h!]
\begin{tabular}{ l | c  c  c  c}
  $$ & I & II & III & IV \\[2pt] 
  \hline 
  \hline                     
  \raisebox{-4pt}{$\hat{c}_0^*$} \hspace{1mm} & \hspace{1mm} \raisebox{-4pt}{$0$} 
  \hspace{1.5mm} & \raisebox{-4pt}{$\frac{1}{3}(2-D)$} \hspace{1.5mm} &  
  \raisebox{-4pt}{$\frac{2}{3}(2-D)$} \hspace{1.5mm} & \raisebox{-4pt}{$2-D$} \\[7pt]
  $\hat{c}_2^*$  \hspace{1mm} & \hspace{1mm} $0$ \hspace{1.5mm} & $\frac{1}{3}(D-2)$ 
  \hspace{1.5mm} &  $\frac{1}{3}(2-D)$  \hspace{1.5mm} & $0$ 
\end{tabular}
\caption{\label{fixed_points1}Different fixed points corresponding to the RG 
equations~\eqref{c0_first} and~\eqref{c2_first}. The dimensionless values are defined as 
$\hat{c}_i^* = 2\ela c_i^*/K_D^{}\Lambda^D$, $i=0,2$. In three dimensions, the Gaussian 
fixed point I is stable and the $SU(3)$ symmetric fixed point IV is unstable. Fixed points II 
and III have both relevant and irrelevant scaling fields.}
\end{table}

\begin{figure}[h!]
\begin{center}
\includegraphics[width=0.375\textwidth]{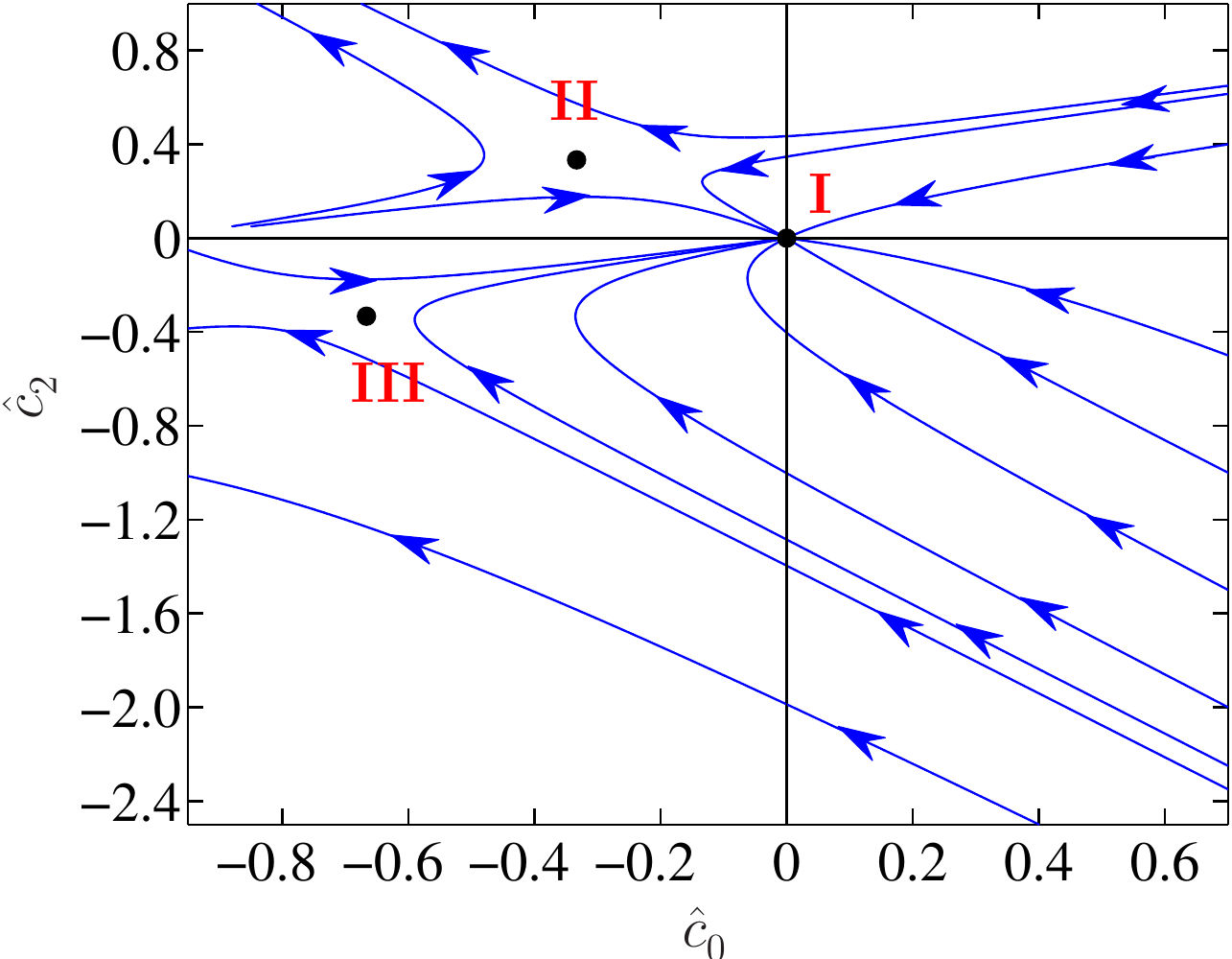}
\caption{\label{rg_flow1} (Color online) Renormalization group flow for a three 
dimensional spin-$1$ Bose gas at zero temperature in the absence of dipole--dipole 
interactions. The dimensionless coupling constants are defined as 
$\hat{c}_i^{} = 2\ela c_i^{}/K_3^{}\Lambda^3$, $i=0,2$. The Gaussian fixed point (I) is 
stable, and RG flows starting from $\hat{c}_0^{}\geq 0$ and small $|\hat{c}_2^{} | $ 
end up to the Gaussian fixed point. Fixed points II and III  are in the physically inaccessible 
region but they still have an effect on the RG flow since they give rise to two runaway flows 
for large $|\hat{c}_2^{}|$.}
\end{center}
\end{figure}

The fixed points corresponding to the RG flow defined by Eqs.~\eqref{c0_first} 
and~\eqref{c2_first} can be determined exactly. The four different fixed points are 
given in Table~\ref{fixed_points1}, and the RG flow for $D=3$ is shown in 
Fig.~\ref{rg_flow1}. The fixed points $\hat{c}_0^*$ and $\hat{c}_2^*$ correspond to 
the dimensionless values $\hat{c}_i^* = 2\ela c_i^*/K_D^{}\Lambda^D$, for $i=0,2$. In the special case $D=2$, the dimensionless quantities are independent of $\Lambda$~\cite{Kolezhuk:2010}. 
The RG flows in Fig.~\ref{rg_flow1} show that similarly to the $D=1$ and $D=2$ cases, 
there are two runaway flows indicating the formation of bound spin singlet pairs (positive 
$c_2^{}$) and ferromagnetic instability (negative $c_2^{}$) where the condensate becomes 
locally fully spin-polarized in the sense that fluctuations in the magnitude of the local 
spin become suppressed. 

\vspace{5mm}

The runaway flow associated with the formation of pair condensate renders the coupling 
$g_0^{}= c_0^{} - 2c_2^{}$ corresponding to the spin singlet channel negative, while the 
coupling $g_2^{}= c_0^{} + c_2^{}$ in the $F=2$ channel remains 
positive. Hence this instability corresponds to formation of spin singlet pairs. 
The second runaway flow where $c_2^{}$ becomes ever more negative renders 
both $g_0^{}$ and $g_2^{}$ negative. We refer to these two instabilities as 
antiferromagnetic (AFM) and ferromagnetic (FM) runaway flow, respectively. 

At mean-field level, stability of a finite cloud against collapse requires the bare coupling 
constants to satisfy $c_0^{}\geq 0$ and $-c_2^{} \leq c_0^{}$~\cite{Yang:2009}. On the 
other hand, the flow diagram in Fig.~\ref{rg_flow1} indicates that many-body corrections 
yield a larger window of coupling constants $\hat{c}_0^{}$ and $\hat{c}_2^{}$ which 
renormalize to the Gaussian fixed point. Although both FM and AFM runaway flows suggest  
that the system becomes unstable, stability can be restored by including the higher order 
terms generated by the RG transformation. Such terms in general are marginal or irrelevant 
in the RG sense, but can nevertheless become important in the regime where RG flow does 
not converge to any fixed point~\cite{Rudnick:1978,Amit:2005}. 

Between the regions corresponding to AFM and FM runaway flows, the gas forms the usual 
spinor condensate. Since interactions tend to renormalize to the Gaussian fixed point, the Bogoliubov mean-field theory of spinor condensates is a valid description of the system in 
this regime (cf.~Ref.~\cite{Fisher:1988}). However, the RG approach used here does not 
provide information about nonuniversal properties such as the possible fragmentation of the condensate for antiferromagnetic interactions~\cite{Ho:2000}. 

\subsection{\label{no_ddi_classical}Thermal regime}

In the thermal regime, we require that $\bt$ in Eq.~\eqref{beta_rg} 
does not flow, i.e., the temperature is kept constant in the RG flow. This implies 
$z=0$, and Eq.~\eqref{scaling_identity} gives $\zeta=(2-D)/2$. From Eq.~\eqref{gamma} 
we observe that any finite initial $\Gamma_0^{}$ flows to zero. Quantum fluctuations 
are thus negligible in this limit and we take $\Gamma \rightarrow 0$ in 
Eqs.~\eqref{mu_rg}--\eqref{c2_rg}. We obtain
\begin{align}
\label{thermal1}
&\Gamma\,\nbe[\bt\Gamma(\ela-\mu)] \rightarrow \frac{1}{\bt(\ela-\mu)}, \\
\label{thermal2}
&\chi_1^{}(\bt,\Gamma) \rightarrow \frac{1}{\bt(\ela-\mu)^2}, \\
\label{thermal3}
&\chi_2^{}(\bt,\Gamma) \rightarrow \frac{1}{\bt(\ela-\mu)^2}, 
\end{align}
and at the critical plane corresponding to $\mu=0$~\cite{rg_note,Aharony:1976} we have 
\begin{align}
\label{c0_second}
&\frac{dc_0^{}}{d\ell} = (4-D)c_0^{} - (5c_0^2 + 2c_2^2)\frac{K_D^{}\Lambda^D}
{\bt\varepsilon_{\Lambda}^2}, \\
\label{c2_second}
&\frac{dc_2^{}}{d\ell} = (4-D)c_2^{} - (3c_2^2 + 6c_0^{}c_2^{})\frac{K_D^{}\Lambda^D}
{\bt\varepsilon_{\Lambda}^2}.
\end{align}
At finite temperatures we define dimensionless coupling constants by   
$\hat{c}_i^{} = \bt\elaa c_i^{}/K_D^{}\Lambda^D$, $i=0,2$. Note that the definition is slightly 
different from the zero-temperature case. Fixed points corresponding to the dimensionless 
coupling constants are shown in Table~\ref{fixed_points2}, where have defined 
$\epsilon = 4-D$. 
 
\vspace{5mm}

\begin{table}[h!]
\begin{tabular}{ l | c  c  c  c}
  $$ & I & II & III & IV \\[2pt] 
  \hline 
  \hline                     
  \raisebox{-4pt}{$\hat{c}_0^*$} \hspace{1mm} & \hspace{1mm} \raisebox{-4pt}{$0$} 
  \hspace{1.5mm} & \raisebox{-4pt}{$0.088\epsilon$} \hspace{1.5mm} &  
  \raisebox{-4pt}{$0.194\epsilon$} \hspace{1.5mm} & \raisebox{-4pt}{$0.2\epsilon$} \\[7pt]
  $\hat{c}_2^*$  \hspace{1mm} & \hspace{1mm} $0$ \hspace{1.5mm} & $0.157\epsilon$ 
  \hspace{1.5mm} &  $-0.054\epsilon$  \hspace{1.5mm} & $0$ 
\end{tabular}
\caption{\label{fixed_points2}Fixed points corresponding to the RG 
equations~\eqref{c0_second} and~\eqref{c2_second}.  The dimensionless values are defined 
as $\hat{c}_i^{*} = \bt\elaa c_i^{*}/K_D^{}\Lambda^D$, $i=0,2$. The Gaussian fixed 
point I is unstable and the $SU(3)$-symmetric fixed point IV is stable for 
$\epsilon >0$ ($D<4$). Fixed points II and III have both relevant and irrelevant 
scaling fields.}
\end{table}

\begin{figure}[h!]
\begin{center}
\includegraphics[width=0.375\textwidth]{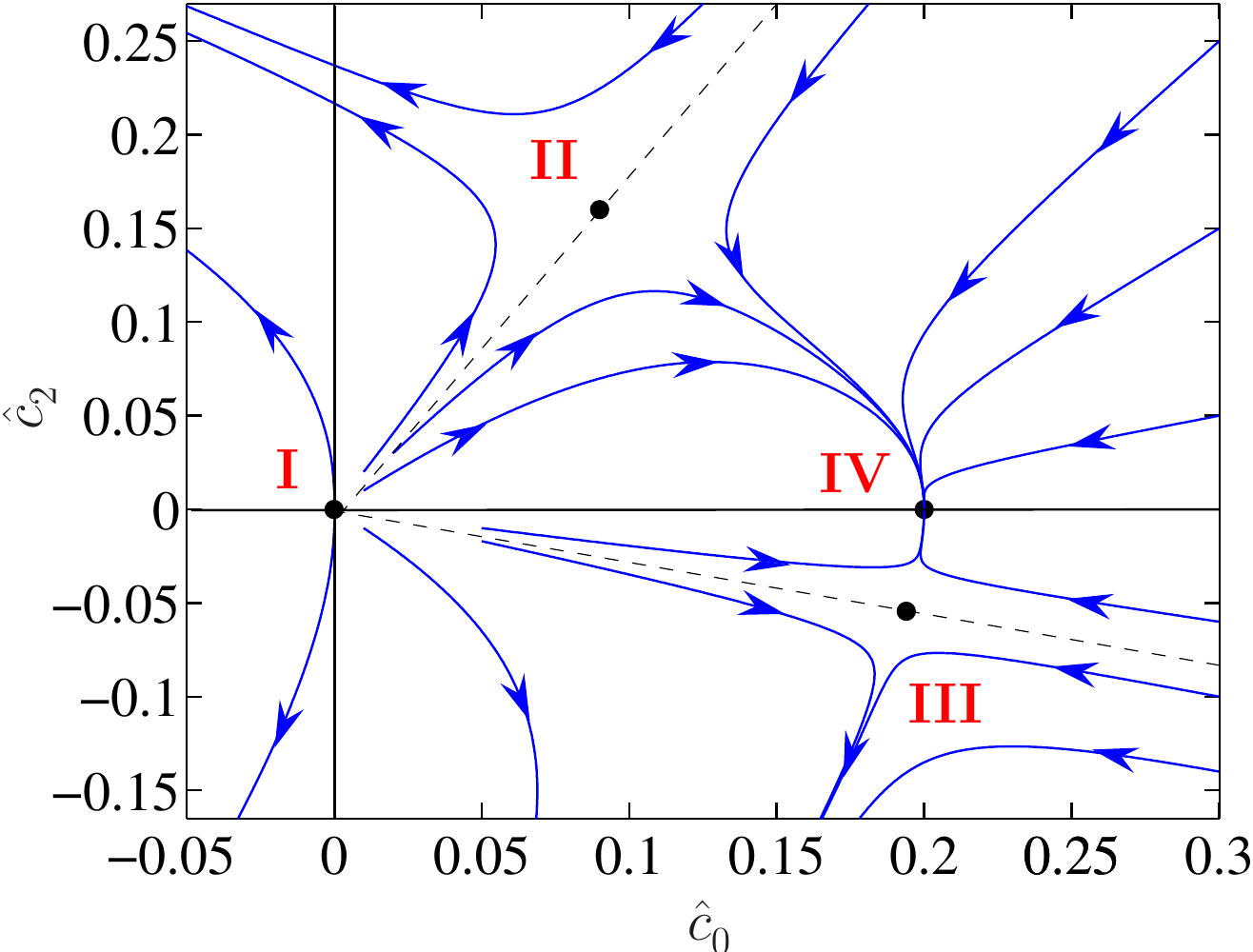}
\caption{\label{rg_flow2} (Color online) Renormalization group flow for a three 
dimensional spin-$1$ Bose gas at fixed temperature. The RG flows correspond 
to the case where dipole--dipole interactions are either absent or take the fixed-point value 
$c_{\mathrm{dd}}^*=0$. The dimensionless coupling constants are defined as 
$\hat{c}_i^{} = \bt\elaa c_i^{}/K_3^{}\Lambda^3$, $i=0,2$. The ferromagnetic condensate 
corresponds to the region with $\hat{c}_2^{}<0$ above the lower dashed line. The region of 
antiferromagnetic condensate is delimited by $\hat{c}_2^{}>0$ and the upper dashed line.}
\end{center}
\end{figure}

To analyze the RG equations in the absence of DDI, we concentrate on the case $D=3$ 
for which the RG flows are depicted in Fig.~\ref{rg_flow2}. The instabilities indicated by 
the runaway flows have the same structure and physical interpretation as in the 
zero-temperature case. An interesting difference to the zero-temperature limit is that the 
ferromagnetic instability corresponding to the runaway flows for large negative $c_2^{}$ 
takes place before the mean-field criterion $c_0^{} \geq 0$ and $-c_2^{} \leq c_0^{}$ 
is violated. Hence, the thermal fluctuations tend to decrease 
the stability of spinor condensates on the ferromagnetic side ($c_2^{} < 0$). 

We note that only the ratio $\hat{c}_2^{}/\hat{c}_0^{}$ is universal (i.e.,~independent of the 
cutoff $\Lambda$), and therefore quantitative comparison of the singlet condensate formation 
in the quantum and thermal regimes depends on $\Lambda$, see Figs.~\ref{rg_flow1} 
and~\ref{rg_flow2}. The values $c_2^{}/c_0^{}$ discussed in 
Sec.~\ref{model} place the bare coupling constants $\hat{c}_0^{}(0)$ and $\hat{c}_2^{}(0)$ for 
$^{23}$Na and $^{87}$Rb into the regime where $\hat{c}_0^{}(\ell)$ and $\hat{c}_2^{}(\ell)$ 
either renormalize to zero (quantum limit) or to the $SU(3)$ symmetric fixed point where 
$\hat{c}_2^{*}=0$ (thermal limit). Since ferromagnetic and antiferromagnetic (polar) 
condensates correspond to different symmetries, they should be separated by a phase 
transition. At mean-field level, we expect the transition to be first order (see also 
Ref.~\cite{Yang:2009}). The RG calculation supports this conclusion in the sense that we do 
not find a critical point separating the two phases, see Fig.~\ref{rg_flow2}.

\vspace{1mm}

At first it may seem surprising that the fixed point IV (or the Gaussian fixed point I at $T=0$) governs the properties of both antiferromagnetic and ferromagnetic condensates. However, according to the exact theorem of 
Ref.~\cite{Eisenberg:2002}, the correct low-energy theory of spin-$1$ bosons should indeed 
correspond to $c_2^{}=0$. The authors of Ref.~\cite{Eisenberg:2002} propose 
that a nonzero $c_2^{}$ in the low-energy theory could arise either from dipole--dipole 
interactions or from the electron transfer between the atoms. 
In Secs.~\ref{ddi_rg_pedestrian} and~\ref{ddi_rg_full} we show that DDI 
is indeed sufficient to give rise to nonzero $c_2^{}$ in the low-energy description of spinor 
Bose gases. Since several important properties of the spinor Bose gases hinge upon the 
presupposition that the spin-dependent coupling is 
nonzero~\cite{Leonhardt:2000,Stoof:2001,AlKhawaja:2001,Savage:2003,Kawaguchi:2008,
Mukerjee:2006,Pietila:2010,Takahashi:2007,Huhtamaki:2010,Pietila:2009}, the analysis in 
Secs.~\ref{ddi_rg_pedestrian} and~\ref{ddi_rg_full} provides justification for this key  
assumption. The two runaway flows do not 
contradict the aforementioned theorem since they correspond to the formation of a spinless 
condensate consisting of either single atoms polarized to the same hyperfine spin state or 
spin singlet pairs.

\section{\label{ddi_rg_pedestrian}RG analysis of dipolar Bose gas}

We calculate contributions arising from DDI using the generic diagrams depicted in 
Figs.~\ref{chempot} and~\ref{couplings}. The contribution from the tadpole diagram 
in Fig.~\ref{chempot}(a) vanishes identically for DDI, and the rainbow diagram in 
Fig.~\ref{chempot}(b) does not contribute to the renormalization of chemical 
potential. However, the rainbow diagram does in principle contribute to the 
anomalous dimension $\eta$ which indicates the importance of renormalization of the 
kinetic energy term in Eq.~\eqref{s0}. At this point, we neglect the renormalization 
$\varepsilon_{\bk}^{}\rightarrow Z_{\eta}^{}\varepsilon_{\bk}^{}$ altogether and set 
$Z_{\eta}^{} = 1$. We will justify this assumption after we have derived the full RG 
equations in Sec.~IV. 

The rainbow diagram also generates a relevant term which in the 
Cartesian basis takes the simple form
\begin{equation}
\label{s0_new}
S_0' = -h_0^{}\int \mathrm{d}^4k\,\,\psi_{\alpha}^*(\vec{k})\,k_{\alpha}^{}k_{\beta}^{}\,
\psi_{\beta}^{}(\vec{k}),\,\,\,\,\al,\bt\in\{x,y,z\}.
\end{equation}
We have introduced here a new coupling constant $h_0^{}$ that is determined by  
the RG equations. At the lowest order, $h_0^{}$ is proportional to 
$c_{\mathrm{dd}}^{}$. We note that the combined contribution of the kinetic energy 
renormalization and $S_0'$ can be written as a squared spin-orbit interaction 
\begin{equation}
S_{\mathrm{so}}^{} = \int \mathrm{d}^4k\,\,\psi_{\alpha}^*(\vec{k})\,(\bk \cdot 
\bm{\mathcal{F}})^2_{\al\bt}\,\psi_{\beta}^{}(\vec{k}).
\end{equation}
The new single-particle term $S_0'$  is allowed since it has the 
same symmetries as the original action given by Eqs.~\eqref{s0} and~\eqref{sint}. Whether 
this term should be included to the original action from the beginning depends on the 
behaviour of $c_{\mathrm{dd}}^{}(\ell)$ in the RG flow and, in particular, on the values of 
$c_{\mathrm{dd}}^{}(\ell)$ at the fixed points. 

\vspace{5mm}

\begin{figure}[h!]
\begin{center}
\includegraphics[width=0.375\textwidth]{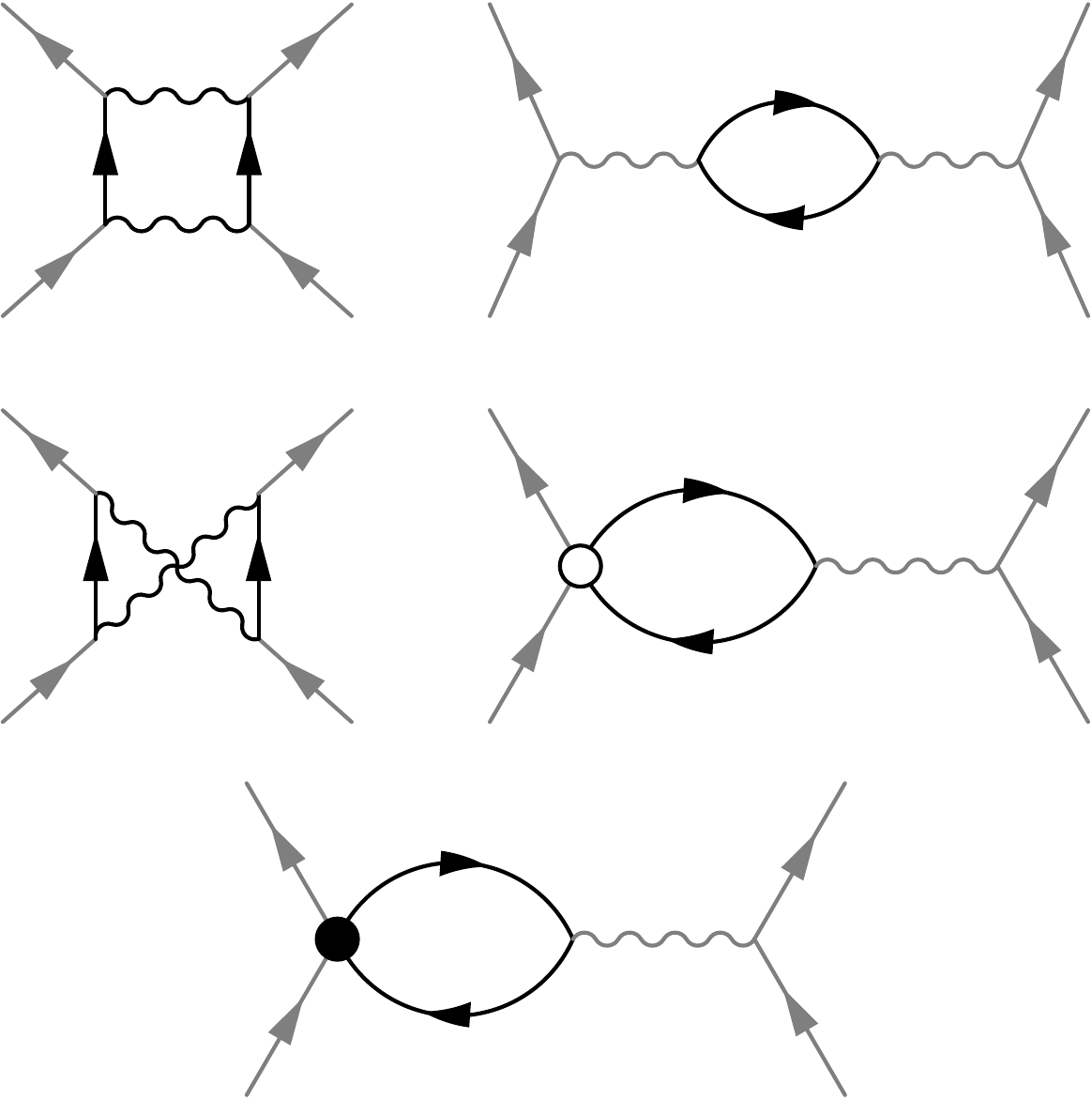}
\caption{\label{dipolar_diagrams} Additional one-loop diagrams giving a nonzero contribution 
to the RG equations in the presence of dipole--dipole interactions.}
\end{center}
\end{figure}

Most of the diagrams in Fig.~\ref{couplings} do not contribute to the renormalization of the 
coupling constants. All the relevant terms are shown in  Fig.~\ref{dipolar_diagrams}, and they 
can be evaluated using essentially the same methods as in Sec.~\ref{no_ddi}. They yield
\begin{align}
\label{ddi1}
&\rd g_s^{} = (\lambda_1^{}c_{\mathrm{dd}}^{2}\chi_1^{} + \lambda_2^{}c_{\mathrm{dd}}^{2}\bt\chi_2^{})K_3^{}
\Lambda^3 \rd\ell \\
\label{ddi2}
&\rd g_d^{} = -(\lambda_3^{}c_{\mathrm{dd}}^{2}\chi_1^{} + \lambda_4^{}c_{\mathrm{dd}}^{2}\bt\chi_2^{})K_3^{}
\Lambda^3 \rd\ell \\
\label{ddi3}
&\rd c_{\mathrm{dd}}^{} = -(\lambda_5^{}c_{\mathrm{dd}}^{2}+2c_{\mathrm{dd}}^{}g_d^{} - 4c_{\mathrm{dd}}^{}g_s^{})K_3^{}\chi_2^{}
\Lambda^3 \rd\ell, 
\end{align}
where $\lambda_1^{}=32\pi^2/45$, $\lambda_2^{} = 272\pi^2/45$, $\lambda_3^{}=256\pi^2/45$, 
$\lambda_4^{}=496\pi^2/45$, and $\lambda_5^{}=8\pi/3$. The complete set of RG equations for 
a dipolar Bose gas consists of contributions from Eqs.~\eqref{ddi1}--\eqref{ddi3} and the 
RG equations~\eqref{beta_rg}--\eqref{c2_rg}. We consider again the 
same limits as in Sec.~\ref{no_ddi}, namely, the quantum limit $T\rightarrow 0$ and the 
thermal limit $\Gamma \rightarrow 0$.

\subsection{Quantum regime}

Using the results of Sec.~\ref{no_ddi_quantum}, we obtain the RG equations for nonzero 
dipole--dipole interaction at the fixed point $\mu^*=0$
\begin{align}
\label{c0_ddi_quantum}
&\frac{\rd c_0^{}}{\rd\ell} = -c_0^{} - (c_0^2 + 2c_2^2 + \gamma_1^{}c_{\mathrm{dd}}^2)
\frac{K_3^{}\Lambda^3}{2\ela}, \\
\label{c2_ddi_quantum}
&\frac{\rd c_2^{}}{\rd\ell} = -c_2^{} + (c_2^2 - 2c_0^{}c_2^{}-\gamma_2^{}c_{\mathrm{dd}}^2)
\frac{K_3^{}\Lambda^3}{2\ela}, \\
\label{cdd_quantum}
&\frac{\rd c_{\mathrm{dd}}^{}}{\rd\ell} = -c_{\mathrm{dd}}^{},
\end{align}
where $\gamma_1^{} = 224\pi^2/45$ and $\gamma_2^{} =32\pi^2/45$. We observe immediately 
that the dipole--dipole coupling constant $c_{\mathrm{dd}}^{}$ renormalizes exponentially to 
zero in the zero-temperature limit, and the new single-particle term Eq.~\eqref{s0_new} remains 
unimportant. The fixed points of the RG flow are shown in Table~\ref{fixed_points1} 
with an addition that $c_{\mathrm{dd}}^{*}=0$ at each fixed point. Furthermore, since 
$c_{\mathrm{dd}}^{}$ appears quadratically in Eqs.~\eqref{c0_ddi_quantum} 
and~\eqref{c2_ddi_quantum}, the stability of the fixed points is not affected by DDI and 
$c_{\mathrm{dd}}^{}$ gives rise to an irrelevant scaling field at these fixed points. Otherwise 
the properties of RG flows are the same as in Sec.~\ref{no_ddi_quantum}.

\subsection{\label{thermal_ddi}Thermal regime}

In the thermal regime, we require again that the temperature does not flow under 
RG, which renders the parameter $\Gamma$ to renormalize exponentially to zero. Using 
Eqs.~\eqref{thermal1}--\eqref{thermal3}, we obtain, for $\mu^*=0$ 
\begin{align}
\label{c0_ddi_first}
&\frac{\rd c_0^{}}{\rd\ell} = c_0^{} - (5c_0^2 + 2c_2^2 + \alpha_0^{}c_{\mathrm{dd}}^2)
\frac{K_3^{}\Lambda^3}{\bt\elaa}, \\
\label{c2_ddi_first}
&\frac{\rd c_2^{}}{\rd\ell} = c_2^{} - (3c_2^2 + 6c_0^{}c_2^{}+\alpha_2^{}c_{\mathrm{dd}}^2)
\frac{K_3^{}\Lambda^3}{\bt\elaa}, \\
\label{cdd_first}
&\frac{\rd c_{\mathrm{dd}}^{}}{\rd\ell} = c_{\mathrm{dd}}^{} - (\alpha_{\mathrm{dd}}^{}
c_{\mathrm{dd}}^2+2c_0^{}c_{\mathrm{dd}}^{} + 
6c_2^{}c_{\mathrm{dd}}^{})\frac{K_3^{}\Lambda^3}{\bt\elaa},
\end{align}
where $\al_0^{} = 448\pi^2/45$, $\al_2^{} = 304\pi^2/45$, and $\al_{\mathrm{dd}}^{} = 8\pi/3$. 
The RG flow gives rise to the four fixed points discussed in Sec.~\ref{no_ddi_classical} 
corresponding to $c_{\mathrm{dd}}^{*}=0$. There are also two additional fixed points given by 
\begin{align}
\label{fp1}
&\hat{c}_0^* = 0.087,\,\,\,\hat{c}_2^* = 0.117,\,\,\,\hat{c}_{\mathrm{dd}}^{*}=0.015, \\
\label{fp2}
&\hat{c}_0^* = 0.088,\,\,\,\hat{c}_2^* = 0.148,\,\,\,\hat{c}_{\mathrm{dd}}^{*}=-0.008.
\end{align}

The dipole--dipole interaction introduces new relevant scaling fields~\cite{scaling_fields} 
for the fixed points I, III, and IV, while the fixed point II has the same instabilities as in the absence 
of DDI. The new dipolar fixed points in Eqs.~\eqref{fp1} and~\eqref{fp2} have both 
relevant and irrelevant scaling fields, and the RG flows in the vicinity of these fixed 
points have properties similar to those of dipolar ferromagnets with spatial 
disorder~\cite{Aharony:1975}. We do not yet dwell on this point since our 
analysis so far has neglected the relevant single-particle term~\eqref{s0_new} generated 
by DDI. Since the dipole--dipole interaction does not renormalize to zero in the thermal 
regime, we have to include the single-particle term in Eq.~\eqref{s0_new} into our 
analysis with an {\it a priori} unknown coupling constant $h_0^{}$ in order to 
properly investigate the dipolar fixed points. 

We analyze the properties of RG flows in the thermal regime more carefully in 
the next section where we consider an extensive model for dipolar Bose gases.  
We also point out that the above 
conclusions hold even if the flow of the chemical potential is taken into account, i.e.,  the 
fixed points and their properties remain qualitatively the same.

\section{\label{ddi_rg_full}Complete RG analysis at finite temperatures}

We analyze here the properties of dipolar spinor Bose gases using the full effective action 
$S_{\mathrm{full}} = S_0^{} + S_0^{'} + S_{\mathrm{int}}^{}$, where $S_0^{}$, $S_0^{'}$, and 
$S_{\mathrm{int}}^{}$ are given by Eqs.~\eqref{s_0}, \eqref{s0_new}, and~\eqref{s_int}. 
Furthermore, we allow renormalization of the kinetic-energy term by redefining 
$\varepsilon_{\bk}^{} = Z_{\eta}^{}\hbar^2\bk^2/2m $ with the initial condition 
$Z_{\eta}^{}(0)=1$. In the RG transformation, the kinetic energy scales as 
$\varepsilon_{\bk}^{} \rightarrow \varepsilon_{\bk}^{} e^{\ln Z_{\eta}^{}(\ell)-2\ell}_{}$, and 
in order to keep the total kinetic energy unchanged, we have to rescale fields as 
$\psi_{\al}^{} \rightarrow \psi_{\al}^{} e^{\xi\ell - \frac{1}{2}\ln Z_{\eta}^{}(\ell)}_{}$.
The quantity $\ln Z_{\eta}^{}(\ell)$ gives rise to anomalous dimension $\eta$ which we 
will discuss in more detail once we have the final RG equations at hand. The anomalous 
dimension changes the scaling relations~\eqref{gamma_scaling}--\eqref{g_scaling} and 
gives $h_0^{}$ a nontrivial scaling  
\begin{subequations}
\label{scaling3}
\begin{align}
\label{gamma_scaling2}
&\Gamma \rightarrow \Gamma\,e^{(d+2\zeta)\ell-Z_{\eta}^{}(\ell)}_{},\\
\label{mu_scaling2}
&\mu \rightarrow \mu\,e^{-2\ell+Z_{\eta}^{}(\ell)}_{}, \\
\label{g_scaling2}
&g \rightarrow g\,e^{-2(\zeta+1)\ell+2Z_{\eta}^{}(\ell)}_{},\\
\label{h0_scaling}
&h_0^{} \rightarrow h_0^{}\,e^{Z_{\eta}^{}(\ell)}_{},
\end{align}
\end{subequations}
where $g = g_d^{}, g_s^{},c_{\mathrm{dd}}^{}$. The appearance of nonzero anomalous 
dimension does not change Eq.~\eqref{scaling_identity}.

Since dipole--dipole interactions were found to be relevant only in the thermal regime 
(Sect.~\ref{thermal_ddi}), we require again that the temperature does not flow. 
We observe that $\Gamma$ does not acquire any renormalization beyond the trivial 
rescaling even in the case of the augmented action $S_{\mathrm{full}}$. Assuming that 
$\ln Z_{\eta}^{}(\ell)$ does not become too large during the RG flow, 
Eq.~\eqref{gamma_scaling2} renders $\Gamma$ to flow to zero. Therefore, both bubble 
diagrams in Fig.~\ref{bubbles} give an equal contribution.

In the presence of the new single-particle operator $S_0^{'}$, the non-interacting 
Green's function becomes non-diagonal and takes the form 
\begin{align}
&\mathcal{G}_{0,\al\bt}(\bk,\omn) = - \frac{\hbar}{-i\hbar\omn + \ebk -\mu}
(\delta_{\al\bt}^{} - \frac{k_{\al}^{}k_{\bt}^{}}{\bk^2}) \notag \\ 
\label{ddi_propagator}
&\hspace{2.3cm} - \frac{\hbar}{-i\hbar\omn + \ebk -\mu -h_0^{}\bk^2} \frac{k_{\al}^{}
k_{\bt}^{}}{\bk^2}. 
\end{align}
We note that in the limit $h_0^{}\rightarrow 0$, Eq.~\eqref{ddi_propagator} reduces to 
Eq.~\eqref{plain_propagator}, and if $h_0^{}\neq 0$, $\mathcal{G}_{0,\al\bt}(\bk,\omn)$ 
is non-singular for $\bk\rightarrow 0$.
A free propagator analogous to that of Eq.~\eqref{ddi_propagator} has been previously 
considered in the context of dipolar magnets, and 
Eq.~\eqref{ddi_propagator} corresponds to the long wave-length limit of the dipolar 
propagator of Ref.~\cite{Aharony:1973}. The difference between the earlier 
studies~\cite{Aharony:1973} and the present work is the itinerant nature of  
magnetism in spinor Bose gases which gives rise to $S_0^{'}$ only through the RG 
transformation. In the theory of classical magnets, single-particle interactions 
similar to $S_0^{'}$ represent the actual dipole--dipole interaction and  
phenomenological quartic terms are taken to be local interactions~\cite{Aharony:1973}. 
In particular, we see later on that the behavior of $h_0^{}$ under the RG flow is different 
between classical dipolar ferromagnets and dipolar spin-$1$ Bose gases, even though 
$h_0^{}$ has formally the same role in both systems.

\begin{figure}[h!]
\begin{center}
\includegraphics[width=0.35\textwidth]{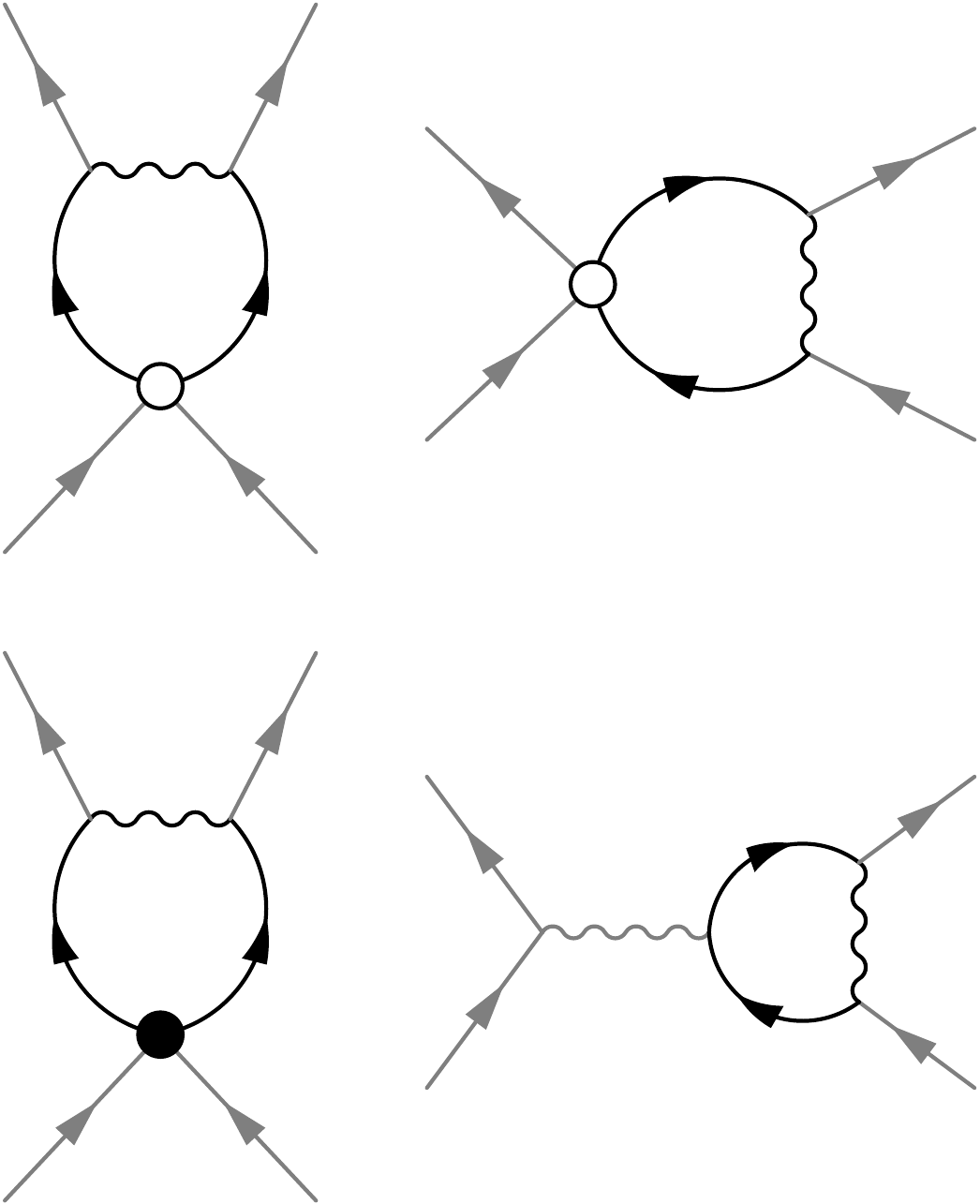}
\caption{\label{full_ddi_diagrams} Additional one-loop diagrams contributing to the RG 
equations due to the nondiagonal part of the Green's function.}
\end{center}
\end{figure}

Since $\mathcal{G}_{0,\al\bt}$ has become nondiagonal, there are new diagrams 
contributing to the renormalization of coupling constants. The new 
diagrams are illustrated in Fig.~\ref{full_ddi_diagrams}. In order to evaluate the angular 
integrals arising from DDI and nondiagonal Green's function, we compute the RG 
equations only up to the linear order in $h_0^{}$. This is a natural approximation, since 
we assume that initially $h_0^{}$ is small if not vanishing. We analyze the accuracy of this 
approximation when we consider the fixed 
points corresponding to the full RG equations. The diagrams in 
Figs.~\ref{no_ddi_diagrams},~\ref{dipolar_diagrams}, and~\ref{full_ddi_diagrams} yield 
the following contributions: 
\begin{align}
& \rd Z_{\eta}^{} = \frac{4m}{15\pi\hbar^2}c_{\mathrm{dd}}^{}(1+5\ho)\chi_{0,1}^{}\Lambda \rd
\ell, \\ 
& \rd h_0^{} = \frac{2}{15\pi}c_{\mathrm{dd}}^{}(3+5\ho)\chi_{0,1}^{}\Lambda \rd\ell, \\
& \rd\mu = \big[-\frac{1}{3\pi^2}(g_s^{}+2g_d^{})(3+\ho) + 
\frac{4}{3\pi}c_{\mathrm{dd}}^{}\ho\big]\chi_{0,1}^{}\Lambda^3 \rd\ell, \\
& \rd g_s^{} = -\frac{1}{2\pi^2}\big[(3g_s^{2}+6g_s^{}g_d^{}-\delta_1c_{\mathrm{dd}}^2)(1+2\ho/3) 
\notag \\
&\qquad\qquad +(\delta_2^{}g_s^{}c_{\mathrm{dd}}^{}+\delta_3^{}g_d^{}c_{\mathrm{dd}}^{} + \delta_4^{}c_{\mathrm{dd}}^2)
\ho\big]\chi_{0,2}^{}\Lambda^3 \rd\ell, \\
& \rd g_d^{} = -\frac{1}{2\pi^2}\big[(5g_d^{2}+4g_d^{}g_s^{}+4g_s^{2}+\delta_5 
c_{\mathrm{dd}}^2)(1+2\ho/3) \notag \\
&\qquad\qquad\qquad\qquad -(\delta_6^{}g_d^{}c_{\mathrm{dd}}^{} + \delta_7^{} 
c_{\mathrm{dd}}^2)\ho\big]\chi_{0,2}^{}\Lambda^3 \rd\ell, \\
& \rd c_{\mathrm{dd}}^{} = -\frac{1}{2\pi^2}\big[(\delta_8^{}c_{\mathrm{dd}}^2+2g_d^{} 
c_{\mathrm{dd}}^{} -4g_s^{}c_{\mathrm{dd}}^{})(1+2\ho/3) \notag \\
&\qquad\qquad\qquad\qquad\qquad\qquad -\delta_9^{}c_{\mathrm{dd}}^{2}\ho\big] 
\chi_{0,2}^{}\Lambda^3 \rd\ell,
\end{align}
where $\ho = h_0^{}\Lambda^2/(\ela-\mu)$ and $\chi_{0,n}^{} = 1/\bt(\ela-\mu)^n$ for 
$n=1,2$. The numerical constants are given by $\delta_1^{}= 304\pi^2/45$, $\delta_2^{}= 
16\pi/9$, $\delta_3^{}= 32\pi/9$, $\delta_4^{}= 128\pi^2/135$, $\delta_5^{}= 752\pi^2/45$, 
$\delta_6^{}= 2\delta_3^{}$, $\delta_7^{}= 3\delta_4^{}$, $\delta_8^{}= 8\pi/3$, and 
$\delta_9^{}= \delta_2^{}$.
In the Zeeman basis, the full RG equations take the form
\begin{align}
\label{eta_final}
& \frac{\rd \ln Z_{\eta}^{}}{\rd\ell} = \frac{8\pi m}{15\hbar^2}c_{\mathrm{dd}}^{}(1+5\ho) 
K_3^{}\Lambda\chi_{0,1}^{}, \\
\label{h0_final}
& \frac{\rd h_0^{}}{\rd\ell} = -\eta h_0^{} + \frac{4\pi}{15}c_{\mathrm{dd}}^{}(3+5\ho)K_3^{}
\Lambda\chi_{0,1}^{}, \\
& \frac{\rd\mu}{\rd\ell} = (2-\eta)\mu - [\frac{2}{3}(c_2^{}+2c_0^{})(3+\ho)  \notag \\ 
&\hspace{4.2cm}- \alpha_0^{}c_{\mathrm{dd}}^{}\ho]K_3^{}\Lambda_{}^3\chi_{0,1}^{}, \\
& \frac{\rd c_0^{}}{\rd\ell} = (1-2\eta)c_0^{} - [(5c_0^{2} +2c_2^{2} + \al_1^{} 
c_{\mathrm{dd}}^2)(1+\frac{2}{3}\ho) \notag \\ 
&\hspace{0.95cm} - (\al_2^{}c_0^{}c_{\mathrm{dd}}^{} + \al_3^{}c_2^{}c_{\mathrm{dd}}^{} +
 \al_4^{}c_{\mathrm{dd}}^{2})\ho]K_3^{}\Lambda_{}^3\chi_{0,2}^{}, \\
& \frac{\rd c_2^{}}{\rd\ell} = (1-2\eta)c_2^{} -  [(3c_2^{2} + 6c_0^{}c_2^{} + \al_5^{}
c_{\mathrm{dd}}^2)(1+\frac{2}{3}\ho) \notag \\
&\hspace{0.95cm} - (\al_6^{}c_2^{}c_{\mathrm{dd}}^{} + \al_7^{}c_0^{}c_{\mathrm{dd}}^{} + 
\al_8^{}c_{\mathrm{dd}}^{2})\ho]K_3^{}\Lambda_{}^3\chi_{0,2}^{}, \\
\label{cdd_final}
& \frac{\rd c_{\mathrm{dd}}^{}}{\rd\ell} = (1-2\eta)c_{\mathrm{dd}}^{} - [(\al_9^{} 
c_{\mathrm{dd}}^2 + 2c_0^{}c_{\mathrm{dd}}^{} + 6c_2^{}c_{\mathrm{dd}}^{}) \notag \\
&\hspace{2.3cm} \times(1+\frac{2}{3}\ho) - \al_{10}^{}c_{\mathrm{dd}}^2\ho]K_3^{} 
\Lambda_{}^3\chi_{0,2}^{},
\end{align} 
where we have defined the anomalous dimension as $\eta = d\ln Z_{\eta}^{}/d\ell$. Comparison 
with Eqs.~\eqref{c0_first},~\eqref{c2_first},~\eqref{c0_second}, and~\eqref{c2_second} 
shows that $\eta$ can be thought of as an effective correction to the spatial dimension of 
the system. Numerical constants $\al_i^{}$ are defined as $\alpha_0^{}=8\pi/3$, 
$\al_{1}^{} = 448\pi^2/45$, $\al_{2}^{} = 32\pi/9$, $\al_{3}^{} = \al_2^{}/2$, 
$\al_{4}^{} = 256\pi^2/135$, $\al_{5}^{} = 304\pi^2/45$, $\al_{6}^{} = \al_3^{}$,  
$\al_{7}^{} = \al_2^{}$, $\al_{8}^{} = \al_4^{}/2$, $\al_{9}^{} = \al_0^{}$, and 
$\al_{10}^{} = \al_3^{}$. The RG equations~\eqref{eta_final}--\eqref{cdd_final} are computed  
up to the linear order $h_0$. 

The fixed points corresponding to RG equations~\eqref{eta_final}--\eqref{cdd_final}
are shown in Table~\ref{fixed_points3} where the dimensionless quantities are defined as 
$\hat{h}^{*}_0 = \Lambda^2 h_0^{*}/\ela$, $\hat{\mu}^{*}_{} = \mu^{*}_{}/\ela$ and 
$\hat{c}_i^* = \beta\elaa c_i^*/K_3^{}\Lambda^3$. Table~\ref{fixed_points3} shows  
that fixed points VII and VIII correspond to relatively large values of $\hat{h}_0^{}$, and 
the original RG equations~\eqref{eta_final}--\eqref{cdd_final} are no longer reliable 
in this region since they were calculated only up to the linear order in $h_0^{}$. 
Linearized RG equations in the vicinity of fixed points V and VI give rise to complex 
eigenvalues. Similar behavior has been 
previously found in the context of dipolar ferromagnets with spatially uncorrelated 
disorder~\cite{Aharony:1975} as well as in systems with long-range-correlated 
disorder~\cite{Weinrib:1983}. We note that the appearance of complex eigenvalues for 
fixed points V and VI could be an artifact of our approximations, and therefore we 
concentrate on the properties of RG flows in the vicinity of fixed points I--IV where 
our calculation should capture the essential physics. 

\vspace{3mm}

\begin{table}[h!]
\begin{tabular}{ l | c  c  c  c  c  c  c  c}
  $$ & I & II & III & IV & V & VI & VII & VIII\\[2pt] 
  \hline 
  \hline  
 
  \raisebox{-4pt}{$\hat{h}_0^*$} \hspace{1mm} & \hspace{1mm} 
  \raisebox{-4pt}{$0$} & 
  \raisebox{-4pt}{$0$} & 
  \raisebox{-4pt}{$0$} & 
  \raisebox{-4pt}{$0$} & 
  \raisebox{-4pt}{$-0.355$} &
  \raisebox{-4pt}{$-0.357$} &
  \raisebox{-4pt}{$1.229$} &
  \raisebox{-4pt}{$1.247$} \\[7pt]  

  $\hat{\mu}_{}^*$ \hspace{1mm} & \hspace{1mm} 
  $0$ &
  $0.250$ & 
  $0.250$ & 
  $0.286$ & 
  $0.282$ & 
  $0.279$ & 
  $0.204$ & 
  $0.124$  \\[7pt]

  $\hat{c}_0^*$ \hspace{1mm} & \hspace{1mm} 
  $0$ &
  $0.050$ &
  $0.109$ &
  $0.102$ &
  $0.066$ &
  $0.070$ &
  $0.028$ &
  $0.034$ \\[7pt]
  
  $\hat{c}_2^*$  \hspace{1mm} & \hspace{1mm} 
  $0$ &
  $0.088$ &
  $-0.031$ &
  $0$ &
  $0.095$ &
  $0.120$ &
  $0.046$ &
  $0.039$ \\[7pt]

  $\hat{c}_{\mathrm{dd}}^*$  \hspace{1mm} & \hspace{1mm} 
  $0$ &
  $0$ &
  $0$ &
  $0$ &
  $0.007$ &
  $-0.009$ &
  $-0.002$ &
  $0.009$ \\[7pt]

  $\eta_{}^*$  \hspace{1mm} & \hspace{1mm} 
  $0$ &
  $0$ &
  $0$ &
  $0$ &
  $-0.012$ &
  $0.015$ &
  $-0.015$ &
  $0.068$

\end{tabular}
\caption{\label{fixed_points3}Dimensionless values of the fixed points corresponding to RG 
equations~\eqref{eta_final}--\eqref{cdd_final}. All fixed points are unstable and 
linearized RG equations possess complex eigenvalues in the vicinity of fixed points 
V,VI, and VIII.}
\end{table}

The Gaussian fixed point I is trivial since it is unstable to every direction in the space of the 
original coupling constants. Fixed points II--IV have certain common features such as the 
existence of one marginal scaling field arising from the combination of $h_0^{}$ and 
$c_{\mathrm{dd}}^{}$. This scaling field reflects the existence of a continuous line of fixed 
points for $c_{\mathrm{dd}}^{*}=0$, corresponding to an arbitrary value of $h_0^{*}$. Since 
$h_0^{}$ was originally  generated by DDI, we have taken $h_0^{*}=0$ for fixed points with 
$c_{\mathrm{dd}}^{*}=0$. Fixed points II--IV also have one scaling field directly proportional to 
$c_{\mathrm{dd}}^{}$. This scaling field is irrelevant for fixed point II and relevant for fixed 
points III and IV.

To further analyze the behavior of RG flows in the case of weak DDI, we simplify the full RG 
equations by taking both $\mu$ and $h_0^{}$ to be critical~\cite{rg_note}. This gives the 
reduced RG equations 
\begin{align}
\label{c0_red}
& \frac{dc_0^{}}{d\ell} = (1-2\eta)c_0^{} - (5c_0^{2} +2c_2^{2} + \al_1^{}c_{\mathrm{dd}}^2)
\frac{K_3^{}\Lambda^3}{\bt\elaa}, \\
\label{c2_red}
& \frac{dc_2^{}}{d\ell} = (1-2\eta)c_2^{} -  (3c_2^{2} + 6c_0^{}c_2^{} + \al_5^{}c_{\mathrm{dd}}^2)
\frac{K_3^{}\Lambda^3}{\bt\elaa}, \\
\label{cdd_red}
& \frac{dc_{\mathrm{dd}}^{}}{d\ell} = (1-2\eta)c_{\mathrm{dd}}^{} - (\al_9^{}c_{\mathrm{dd}}^2 + 2c_0^{}c_{\mathrm{dd}}^{} + 
6c_2^{}c_{\mathrm{dd}}^{})\frac{K_3^{}\Lambda^3}{\bt\elaa},
\end{align}
where the anomalous dimension is given by $\eta = \frac{8\pi m}{15\hbar^2} 
c_{\mathrm{dd}}^{}K_3^{}/\bt\Lambda\ela$. Apart from the anomalous dimension, the reduced RG 
equations are the same as Eqs.~\eqref{c0_ddi_first}--\eqref{cdd_first}. 

The reduced RG equations~\eqref{c0_red}--\eqref{cdd_red} can be used to justify the 
approximation of Sec.~\ref{ddi_rg_pedestrian}, where the anomalous dimension was 
neglected altogether. Equations~\eqref{c0_red}--\eqref{cdd_red} have four fixed points 
I--IV shown in Table~\ref{fixed_points2} corresponding to vanishing DDI (in 3D, we take 
$\epsilon=1$). In addition, there are two other non-trivial fixed points  
\begin{align}
\label{fp3}
&\hat{c}_0^* = 0.085,\,\,\,\hat{c}_2^* = 0.114,\,\,\,\hat{c}_{\mathrm{dd}}^{*}=0.015, \\
\label{fp4}
&\hat{c}_0^* = 0.090,\,\,\,\hat{c}_2^* = 0.150,\,\,\,\hat{c}_{\mathrm{dd}}^{*}=-0.008.
\end{align}
Comparing Eqs.~\eqref{fp1} and~\eqref{fp2} to Eqs.~\eqref{fp3} and~\eqref{fp4}, one 
observes that the effect of anomalous dimension is negligible. Furthermore, 
Table~\ref{fixed_points3} shows that the values of $\eta$ at fixed points are small 
compared to unity and hence the anomalous dimension has only a small effect on the 
fixed point structure of the full RG equations~\eqref{eta_final}--\eqref{cdd_final}.

In the vicinity of fixed points I--IV, linearized RG equations again give rise to one scaling field 
that is directly proportional to the dipole--dipole interaction. In the case of fixed points I, III, and 
IV, this scaling field is relevant and DDI introduces an additional instability. To quantify this 
instability, we define crossover exponents~\cite{Ma:1976}  
$\phi_i^{} = \lambda_i^{}/\lambda_{\scriptscriptstyle\mathrm{DDI}}^{}$, where 
$\lambda_{\scriptscriptstyle\mathrm{DDI}}^{}$ is the eigenvalue corresponding to the 
DDI-induced scaling field and $\lambda_{1,2}^{}$ are the two remaining eigenvalues 
corresponding to the flows depicted in Fig.~\ref{rg_flow2}. The crossover exponent 
$\phi_i^{}$ indicates the relative importance of a given scaling field $s_i^{}$ with respect to 
the DDI induced instability~\cite{scaling_fields}. When the absolute value of the crossover 
exponent becomes smaller than unity, the DDI-dependent instability dominates the RG flow with respect to $s_i^{}$. 

\begin{table}[h!]
\begin{tabular}{ l | c  c  c  c}
  $$ & I & II & III & IV \\[2pt] 
  \hline 
  \hline                     
  \raisebox{-4pt}{$\phi_1^{}$} \hspace{1mm} & \hspace{1mm} \raisebox{-4pt}{$1$} 
  \hspace{1.5mm} & \raisebox{-4pt}{$5.498$} \hspace{1.5mm} &  
  \raisebox{-4pt}{$-1.069$} \hspace{1.5mm} & \raisebox{-4pt}{$-1.667$} \\[7pt]
  $\phi_2^{}$  \hspace{1mm} & \hspace{1mm} $1$ \hspace{1.5mm} & $-8.464$ 
  \hspace{1.5mm} &  $0.235$  \hspace{1.5mm} & $-0.333$ 
\end{tabular}
\caption{\label{crossover_exponents} Crossover exponents corresponding to fixed points 
I--IV in Table~\ref{fixed_points2}. For the fixed point II, we have replaced $\lambda_{\scriptscriptstyle\mathrm{DDI}}^{}$ with $|\lambda_{\scriptscriptstyle\mathrm{DDI}}^{}|$ since the 
DDI scaling field is irrelevant at fixed point II.}
\end{table}

\begin{figure}[h!]
\begin{center}
\includegraphics[width=0.48\textwidth]{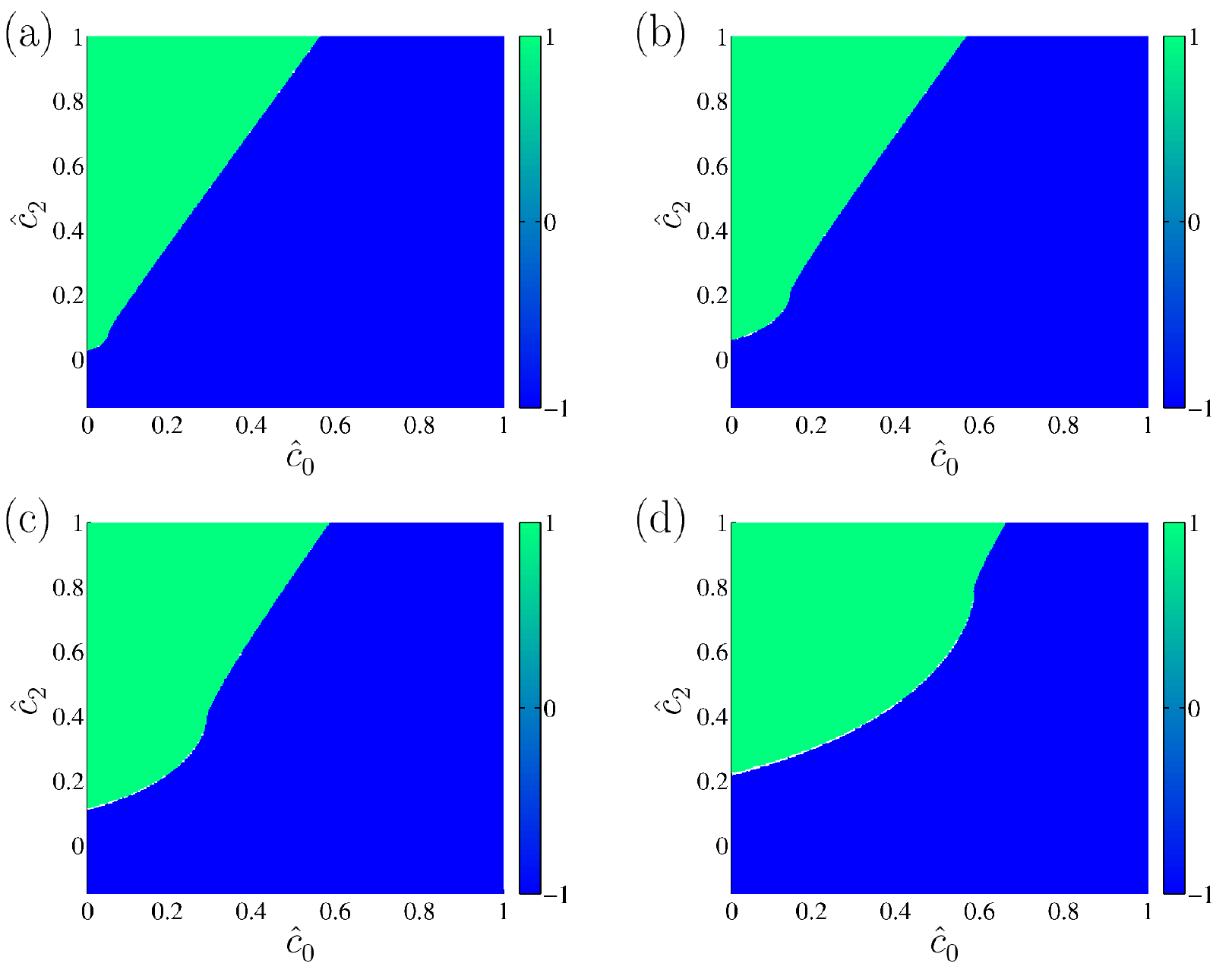}
\caption{\label{c2_flow} (Color online) Sign of the dimensionless coupling constant 
$\hat{c}_2^{}(\ell_c)$ for each initial point 
$(\hat{c}_0^{},\hat{c}_2^{},\hat{c}_{\mathrm{dd}}^{})$. The dimensionless coupling constants 
are defined as $\hat{c}_i^{} = \bt\elaa c_i^{}/K_D^{}\Lambda^D$, $i=0,2,\mathrm{dd}$.
Reduced RG equations~\eqref{c0_red}--\eqref{cdd_red} are integrated up to $\ell_c$, 
where $\ell_c$ is determined by the condition $|\hat{c}_{2}^{}(\ell_c)| = 10$. The 
initial values of $\hat{c}_{\mathrm{dd}}^{}$ are (a) $\hat{c}_{\mathrm{dd}}^{}(0)=0.01$, 
(b) $\hat{c}_{\mathrm{dd}}^{}(0)=0.025$, (c) $\hat{c}_{\mathrm{dd}}^{}(0)=0.05$, and (d) 
$\hat{c}_{\mathrm{dd}}^{}(0)=0.1$.}
\end{center}
\end{figure}

\begin{figure}[h!]
\begin{center}
\includegraphics[width=0.48\textwidth]{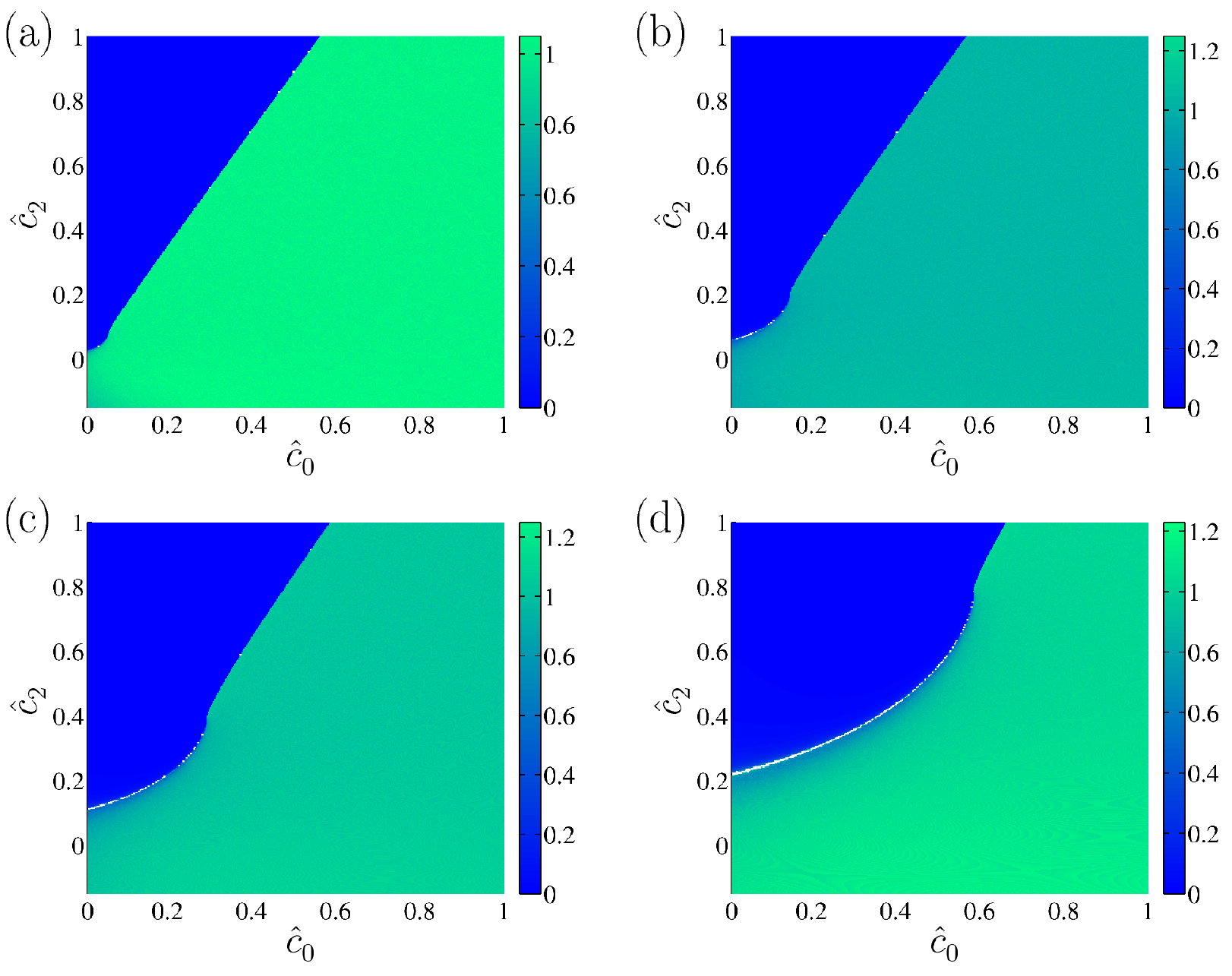}
\caption{\label{cdd_flow} (Color online) Value of the dimensionless coupling constant 
$\hat{c}_{\mathrm{dd}}^{}(\ell_c)$ for different initial points 
$(\hat{c}_0^{},\hat{c}_2^{},\hat{c}_{\mathrm{dd}}^{})$. The reduced RG 
equations are integrated up to $\ell_c$, where $\ell_c$ is determined by the condition 
$|\hat{c}_{2}^{}(\ell_c)| = 10$. The initial values of $\hat{c}_{\mathrm{dd}}^{}$ are the same as 
in Fig.~\ref{c2_flow}, namely (a) $\hat{c}_{\mathrm{dd}}^{}(0)=0.01$, (b) 
$\hat{c}_{\mathrm{dd}}^{}(0)=0.025$, (c) $\hat{c}_{\mathrm{dd}}^{}(0)=0.05$, and (d) 
$\hat{c}_{\mathrm{dd}}^{}(0)=0.1$.}
\end{center}
\end{figure}

The crossover exponents are shown in Table~\ref{crossover_exponents} and they indicate 
that the properties of fixed point II are largely unaffected by the dipole--dipole interaction. 
On the other hand, fixed points I, II, and IV are susceptible to the runaway flows induced by DDI. 
To explore these runaway flows, we integrate the RG equations numerically in the vicinity 
of fixed points I--IV. We take $c_{dd}(0)$ to be small and positive, which is the physically 
relevant case. We find that the system exhibits the two runaway flows discussed in 
Sect.~\ref{no_ddi} (see~Fig~\ref{rg_flow2}). However, the mean-field regime governed by 
the fixed point IV becomes unstable and RG flows starting from this region 
correspond to the ferromagnetic runaway flow. The dipole--dipole interaction tends to 
increase under the ferromagnetic runaway flow, whereas under the antiferromagnetic 
runaway flow DDI renormalizes to zero. This is demonstrated in Figs.~\ref{c2_flow} 
and~\ref{cdd_flow} where we show the sign of $c_2^{}$ (Fig.~\ref{c2_flow}) and the 
magnitude of $c_{\mathrm{dd}}^{}$ (Fig.~\ref{cdd_flow}) in the asymptotic limit of RG flows.
For each initial point $(\hat{c}_0^{},\hat{c}_2^{},\hat{c}_{\mathrm{dd}}^{})$, we integrate the 
RG equations up to value $\ell_c$, where $\ell_c$ is given by the 
condition $|\hat{c}_{2}^{}(\ell_c)| = 10$. Since both $\hat{c}_0^{}$ and $\hat{c}_2^{}$ diverge 
along the runaway flows, the precise value of  $|\hat{c}_{2}^{}(\ell_c)|$ which determines 
$\ell_c$ is unimportant. It only needs to be large enough to illustrate the general tendency 
associated with the two runaway flows: for AFM runaway flow DDI renormalizes to zero 
whereas for FM runaway flow DDI tends to grow.

Figures~\ref{c2_flow} and~\ref{cdd_flow} demonstrate the existence of a phase transition 
separating an antiferromagnetic spin singlet condensate and a ferromagnetic condensate 
with anisotropic dipole--dipole interactions and suppressed fluctuations in the magnitude of 
local spin (dipolar ferromagnetic condensate). The dipole--dipole interaction favors spatial 
modulation in the local magnetization and in contrast to Sect.~\ref{no_ddi_classical}, the 
equilibrium phase corresponding to the ferromagnetic runaway does not feature all atoms 
in the same hyperfine spin state. Furthermore, even weak DDI renders antiferromagnetic 
and ferromagnetic condensates unstable towards formation of dipolar ferromagnetic 
condensate.

To verify that the anisotropic kinetic energy term represented by $h_0^{}$ does not change 
the previous conclusions, we lift the assumption $h_0^{}=0$ and integrate RG 
equations~\eqref{eta_final}--\eqref{cdd_final} in the critical region corresponding $\mu=0$. We  
take $h_0^{}$ initially zero since the original action does not contain the term in 
Eq.~\eqref{s0_new}. We obtain RG flows identical to those depicted in Fig.~\ref{c2_flow} 
and~\ref{cdd_flow}. For 
the physically relevant case $c_{\mathrm{dd}}^{}(0) > 0$, we find that $h_0^{}$ 
grows under both runaway flows. Under the antiferromagnetic runaway flow, the growth of 
$h_0^{}$ is somewhat slower than in the case of ferromagnetic runaway flow. 

We conclude this section by analysing the stability of the two additional fixed points given in 
Eqs.~\eqref{fp3} and~\eqref{fp4}. As discussed in Sec.~\ref{thermal_ddi}, the RG flow 
in the vicinity of these new fixed points has somewhat unconventional properties that are 
similar to those discussed in Refs~\cite{Aharony:1975,Weinrib:1983}. When the RG 
equations~\eqref{c0_red}--\eqref{cdd_red} are linearized in the vicinity of fixed points, the eigenvalues of the resulting matrix consist of one real eigenvalue and a 
pair of complex-conjugated eigenvalues in the case of physically relevant fixed 
point in Eq.~\eqref{fp3}. The real parts of eigenvalues determine the stability of the fixed 
point~\cite{Weinrib:1983}, and we find that the fixed point in Eq.~\eqref{fp3} is unstable and 
practically unattainable since the complex eigenvalues have positive real parts. Furthermore, 
RG flows near the fixed point in Eq.~\eqref{fp3} are not markedly different from the RG flows 
near fixed points I--IV. We find that RG flows starting from positive $c_{\mathrm{dd}}^{}$ remain 
positive and since  the bare value of DDI coupling $c_{\mathrm{dd}}^{}(0) = 
\mu_0^{}\mu_{\scriptscriptstyle\mathrm{B}}^{2}g_{\scriptscriptstyle\mathrm{F}}^{2}/4\pi$ is 
positive, the fixed point in Eq.~\eqref{fp4} corresponding to negative DDI coupling is not 
relevant for physical systems. For completeness, we note that linearized RG equations for the 
fixed point in Eq.~\eqref{fp4} give rise to real eigenvalues, two of which are positive.

\section{Effects associated with the anisotropic dispersion}
To analyze the properties of the anisotropic kinetic energy term in Eq.~\eqref{s0_new},  
we restrict to the noninteracting limit and consider only the single-particle part $S_0 + S_0'$. 
Since the effects associated with dipole--dipole interactions become important only at 
finite temperatures, we neglect all nonzero Matsubara frequencies in the correlation function 
$G_{0,\al\bt}=\la\psi_{\al}^*\psi_{\bt}^{}\ra$ and obtain 
\be
G_{0,\al\bt}(\bk) = \frac{1}{\ebk-\mu}\big(\delta_{\al\bt}^{} - \frac{k_\al^{}k_\bt^{}}{\bk^2}
\big) + \frac{1}{\ebk-\mu-h_0^{}\bk^2}\frac{k_\al^{}k_\bt^{}}{\bk^2}.
\ee
The anisotropic dispersion does not affect local spin order, but can 
nevertheless give rise to nematic order described by a tensor order parameter
\be
(\mathcal{Q}_s)_{\al\bt}^{} = \frac{1}{2}(\psi_{\al}^{*}\psi_{\bt}^{} + 
\psi_{\bt}^{*}\psi_{\al}^{}).
\ee
Since we consider a uniform system with constant total density, our definition of 
$\mathcal{Q}_s$ is analogous to that of Ref.~\cite{Mueller:2004}.

In the non-interacting limit we have 
$\langle \mathcal{Q}_s(\bk) \rangle = G_{0,\al\bt}^{}(\bk)$. The nematic order is 
associated with the eigenvalues and eigenvectors of $\langle \mathcal{Q}_s \rangle$. 
Following Ref.~\cite{Mueller:2004}, we define the nematic director $\un$ to be the 
eigenvector corresponding to the largest eigenvalue of $\langle \mathcal{Q}_s \rangle$. 
Since $\langle \mathcal{Q}_s\rangle$ has to be positive semidefinite, we impose a 
condition $|h_0^{}|< \hbar^2/2m$. This condition is also physically motivated since initially 
in the RG calculation, we assumed $h_0^{}$ to be small compared to kinetic energy. 
Without losing generality we may also take $\mu=0$. For  $h_0^{}>0$ we obtain 
$\un(\bk) = \hat{\bk}$, corresponding to a hedgehog (monopole) in the momentum space. 
In the case $h_0^{}<0$, there are two linearly independent nematic directors which can be 
taken to be $\un_1^{}(\bk) = (-k_z^{},0,k_x^{})/k_{xz}^{}$ and 
$\un_2^{}(\bk) = (-k_y^{},k_x^{},0)/k_{xy}^{}$, where $k_{x\al}^{} = \sqrt{k_x^{2}+k_\al^{2}}$. 
Such nematic directors correspond to vortices in momentum space.

One of the experimental manifestations of dipolar interactions in spin-$1$ Bose gases 
is periodic spatial modulation in the local magnetization~\cite{Vengalattore:2010}. Since 
we have considered only the noninteracting case in this section, 
the local spin order vanishes and our analysis cannot be directly compared 
with the experiment of Ref.~\cite{Vengalattore:2010}. To study the nematic order associated 
with $h_0^{}$ in real space, we Fourier transform $G_{0,\al\bt}$. The nematic director $\un(\br)$ 
corresponds to the hedgehog for $h_0^{} < 0$ and the two vortex solutions for 
$h_0^{} > 0$. Note that the hedgehog in real space corresponds to a vortex in momentum 
space and vice versa. The nematic ordering and the corresponding textures can in principle 
be measured in real space utilizing the atomic birefringence~\cite{Carusotto:2004,Higbie:2005}.

\section{Discussion}

In this work, we have analyzed the properties of dipolar spin-$1$ Bose 
gases using momentum shell renormalization group and taking into account all one-loop 
diagrams.  In the absence of magnetic dipole--dipole interactions, our RG analysis  
complements the previous RG studies of low-dimensional spinor Bose gases 
at $T=0$~\cite{Kolezhuk:2010} as well as the phenomenological analysis of 
Ref.~\cite{Yang:2009}. Similarly to Ref.~\cite{Kolezhuk:2010}, we found two runaway 
flows corresponding to the formation of either a condensate of spin singlet pairs or a 
fully spin-polarized scalar condensate. The absence of stable fixed points in the regions  
corresponding to runaway flows is sometimes a manifestation of a first-order 
transition~\cite{Amit:2005,Rudnick:1978}. In the case of antiferromagnetic runaway flow, 
it would be interesting to study if stable fixed points arise when an additional field representing 
the spin singlet pairs is introduced~\cite{Kolezhuk:2010}. We believe that the possible 
first-order transition associated with the ferromagnetic runaway flow could be akin to the 
fluctuation-induced first-order transition in type I superconductors~\cite{Halperin:1974}. 
For spinor Bose gases, the analogue of an intrinsic fluctuating magnetic field is given by 
the fluctuating Berry phase associated with the local magnetization~\cite{Ho:1996}.

In the zero-temperature limit, we found that the dipole--dipole interaction renormalizes to zero 
and does not induce any additional instabilities. At finite temperatures, we analyzed the limit 
where thermal fluctuations dominate quantum fluctuations. We found that the pair 
condensate is unaffected by the dipole--dipole interactions which eventually 
renormalize to zero. On the other hand, both antiferromagnetic and ferromagnetic 
condensates become unstable and the system exhibits an instability similar to the 
ferromagnetic runaway flow in the absence of DDI. In principle the magnetic 
dipole--dipole interaction can be transformed to an external vector potential 
such that the local spin of the gas $\bm{\mathcal{S}}$ couples linearly to the 
curl of the vector potential~\cite{Belitz:2010}. This transformation gives rise to an 
alternative RG scheme which could provide further insight to the runaway flow induced by the 
DDI and its connection to the potential first-order transition. Also, the role of higher order terms 
beyond the one-loop approximation should be explored and one possible route to accomplish 
this task could be the functional renormalization group, from which the current RG equations 
arise in principle as an approximation~\cite{Wetterich:1991,Andersen:2004}. 

Since the lifetime of ultracold atomic gases is limited and the spin--spin interactions are 
relatively weak, it is not clear to what extent the current 
experiments~\cite{Vengalattore:2008,Vengalattore:2010} are able to explore the true 
thermal equilibrium of the system. However, even if the experimentally attainable 
physics of spinor Bose gases eventually turns out to be inherently out of equilibrium, 
understanding the corresponding equilibrium systems is still a prerequisite for the 
exploration of the non-equilibrium situation. The experimentally relevant atomic species 
$^{23}$Na and $^{87}$Rb give rise to bare coupling constants that belong to the regions of 
antiferromagnetic and ferromagnetic condensates, respectively. In the absence of dipole--dipole 
interactions, the critical properties of both of these condensates are determined by the 
$SU(3)$-symmetric fixed point discussed in Section~\ref{no_ddi_classical}. When dipole--dipole 
interactions are taken into account, ferromagnetic and antiferromagnetic condensates become 
in principle unstable and the true equilibrium is determined by the ferromagnetic runaway flow. 
However, the lifetime of atomic gases can limit the possibilities of observing this crossover 
from the critical behavior determined by the non-dipolar fixed points of 
Sec.~\ref{no_ddi_classical} and~\ref{ddi_rg_full} to the thermodynamic 
equilibrium determined by DDI. 

Recently observed optical Feshbach resonances~\cite{Hamley:2009} as well as the proposed microwave induced Feshbach resonances~\cite{Papoular:2010} provide in principle means to fully explore the phase diagrams studied in this work. Alternatively, the phase diagrams in the absence of DDI could also be realized in the molecular superfluid phase of 
$p$-wave resonant Bose gases~\cite{Radzihovsky:2009}. 

\begin{acknowledgements}

We would like to thank E.~Demler, D.~Podolsky, and P.~Kuopanportti for insightful comments.
This work was supported by the Academy of Finland (VP and MM), Emil Aaltonen Foundation 
(VP and MM), and  Harvard--MIT CUA (VP).

\end{acknowledgements}

\bibliography{manu} 

\begin{thebibliography}{70}
\bibitem [{sca()}]{scaling_fields}%
  \BibitemOpen
  \href@noop {} {}\bibinfo {note} {The scaling fields are defined as
  projections of the original coupling constants $(c_0^{},c_2^{},c_{dd}^{})$ on
  the left-eigenvectors of the matrix corresponding to the linearized RG
  equations.}\BibitemShut {Stop}%
\end{thebibliography}

\end{document}